\documentclass{svmult}

\usepackage{newtxtext}
\usepackage{newtxmath}
\usepackage{makeidx}
\usepackage{graphicx}
\usepackage{multicol}
\usepackage{footmisc}
\usepackage[T1]{fontenc}
\usepackage{url}

\title*{Dark Matter}
\author{Sergio Luigi Cacciatori, Vittorio Gorini, Federico Re}

\institute{Sergio Luigi Cacciatori \at Department of Science and High Technology, Università dell'Insubria, Via Valleggio 11, Como, 22100, Italy. \\ \email{sergio.cacciatori@uninsubria.it}
\and Vittorio Gorini \at Department of Science and High Technology, Università dell'Insubria, Via Valleggio 11, Como, 22100, Italy. \\ \email{vittorio.gorini@gmail.com}
\and Federico Re \at Department of Physics Giuseppe Occhialini, Università di Milano-Bicocca, Piazza della Scienza 3, 20126, Milano, Italy. \\ \email{federico.re@unimib.it}}

\begin{document}

\maketitle

\begin{abstract}
    {In this article we address the mystery of dark matter. We expound the various evidences, astrophysical and cosmological, leading to hypothesize the existence of an invisible form of matter, whose attempts at detecting it have so far all failed.\\
    We also discuss some alternative suggestions that replace the hypothesis of exotic matter with the assumption of modifications of the gravitational dynamics. For each of the various proposals we also discuss the strong and weak points.\\[0.3cm]
    {\it ``This is a preprint of the following chapter: S. L. Cacciatori, V. Gorini, F. Re, Dark Matter, published in New Frontiers in Science in the Era of AI, edited by M. Streit-Bianchi \& V. Gorini, 2024, Springer reproduced with permission of Springer Nature Switzerland AG. The final authenticated version is available online at: \url{http://dx.doi.org/10.1007/978-3-031-61187-2}''.} See \cite{url}.
    }
\end{abstract}

 \vspace{1cm}
    {\bf Key words}: Dark Matter; Rotation Curves; Gravitational Lensing; MACHO; WIMP; Axions; MOND.

\vspace{1cm}

\section{Basic concepts}\label{DM:2.1}

\subsection{Dark matter is not antimatter}\label{DM:2.1.1}

Confusing dark matter with antimatter is unfortunately a common mistake among people who have some interest in physics, but are not specialists. This confusion is understandable, since antimatter is not something that one sees in daily life. Like dark matter, antimatter seems to many a mysterious, exotic kind of matter. However, they are two very different things.

Antimatter is very well known in physics, by almost a century. For example, the electron has its antimatter counterpart, the positron, well known for instance from the functional imaging technique of positron emission tomography (PET). The positron has the same mass and spin as the electron, but opposite electric charge. Similarly, the proton has its associate antiparticle, the antiproton; the neutron the antineutron; and so on. Nowadays, we often produce antimatter in laboratories and we have a well-established theory (the so called Standard Model of Particles) that accurately describes the interactions among all particles, the matter and the antimatter ones. When a particle of matter gets close to its antimatter counterpart, the two particles annihilate each other and transform their energy into a burst of electromagnetic radiation (gamma rays). For example, PET is based on the detection of the gamma rays generated by the annihilation of pairs of positrons and ordinary electrons of organic tissue, the positrons being produced by a suitable radioactive source. In our universe, the percentage of antimatter is tiny compared to the amount of ordinary matter. This is fortunate. Indeed, if the two amounts were the same, the two components would annihilate each other and the universe would reduce itself to pure radiation.

On the other hand, dark matter \emph{is not known at all}. We do not know what it is made of. We do not know if we need a new theory to describe it. We do not even know if it really exists. Most of the scientific community believes that it does exist. Indeed, although it has never been directly detected so far, we have many indirect evidences of its existence, and these evidences concern independently different phenomena. We will describe them in detail in Section \ref{DM evidencies}.

Metaphorically speaking, we may say that, for a physicists, antimatter is like an often encountered friend. Whereas dark matter is like a stranger who moves some objects around us, but whom we never see.

\section{The many evidences of the dark matter}
\label{DM evidencies}
\label{DM:2.2}

Concerning dark energy, we have seen (see the Chapter on dark energy) that the belief on the existence on this form of energy is based on the purported accelerated expansion of the universe, which is testified by just two distinct evidences: the observation of the redshift of type $Ia$ supernovae and the rate at which structures started to form from the original inhomogeneities caused by the baryon acoustic oscillations (BAO) in the early universe, prior to the decoupling of radiation from matter.
If these observations were to be disproved or otherwise explained in a different way (see Section 2 of the Chapter on dark energy), then we would have no more motivation in believing in dark energy. For dark matter the situation is different. \\
We know several phenomena and observations that suggest the existence of some kind of dark matter. They are all of gravitational nature. All these cases have in common the fact that one is able to estimate the gravitational attraction of a given system of masses, only to find out that the amount of mass that we can see (the visible mass) is not enough to justify the measured gravitational acceleration. Therefore these evidences concern systems displaying "missing mass". These evidences are many and independent one another. This means that an alternative explanation of one of them would not necessarily lead to the explanation of the others as well, that could thus remain a mystery. This situation renders the existence of dark matter more solid than that of dark energy.

\subsection{Astrophysical evidences for dark matter}\label{DM:2.2.1}

Given the great number of evidences for the existence of dark matter, it is convenient to address them with some order. We start from the phenomena of astrophysical relevance (recall that the astrophysical systems are those bound by gravitational forces, hence not larger than galactic clusters and superclusters).

\subsubsection{Galaxy rotation curves}\label{DM:2.2.1.1}

Disc galaxies - like ours, the Milky Way - bear some similarities to our Solar System. The latter is formed by the Sun and its eight planets, plus many more smaller bodies. Save for the outer Oort Cloud, most of these objects orbit the Sun on essentially the same plane. A galaxy is made up of many bodies (tens or hundreds billion stars). Some galaxies have the form of a disc because their stars have the tendency to revolve on a common rotation plane. In the case of the Solar System, the farther a planet is from the Sun, the smaller is its revolving speed. This is described by Kepler's Third Law, according to which the revolving speed of the planet is inversely proportional to the square root of its distance from the star. Indeed, the attractive force exerted by the Sun decreases as the square of the distance of the planet from the star. \\
However, Kepler's Third Law does not hold in a disc galaxy. The reason for this is that, whereas in a planetary system almost all the mass resides in the star, so that the latter is practically the only source of gravitation, a galaxy is instead an extended source whose mass is by no means concentrated in its centre. The farther is a star from the galactic centre, the greater is the mass contained inside its orbit, and this cumulative mass may more or less compensate the Keplerian decrease of the speed. One may draw a plot of the average value of the speed of rotation for any given radius of the orbit. In this way one obtains the so-called \emph{galaxy rotation curve}, from which it is possible to deduce the distribution of the gravitational mass inside the galaxy. \\
Now, it is right here that the problem arises: we can calculate the mass of the galaxy by counting its stars and by taking into account its gas and dust content and we find that it is not sufficient to explain the observed rotation curve. This fact is explained by calling upon the existence of some extra mass that is \emph{invisible}: a dark matter.

\begin{figure}
\label{Figure1}
	\begin{minipage}[b]{1.0\linewidth}
		\centering
		\includegraphics[width=8 cm]{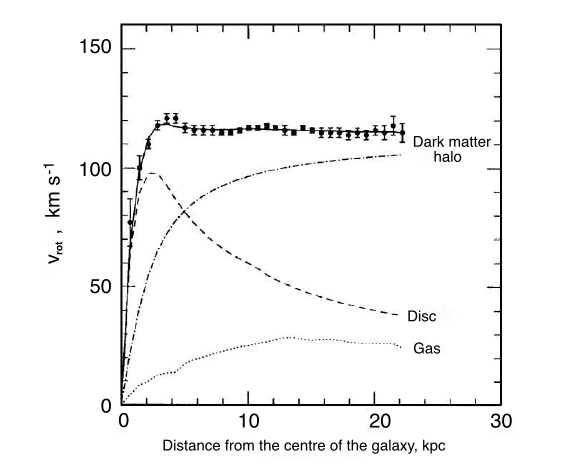}
		\caption{\label{1.1} Example of a galaxy rotation curve. The points with the error bars provide a flat graph up to well the periphery of the galaxy, in a region where there are still a few scanty stars orbiting, as well as gas and dust. This curve cannot be explained solely by the visible matter of the \emph{disc}, that would provide a decreasing rotation curve (the dashed one). This is the reason why one assumes the additional presence of a dark matter \emph{halo}. (Adapted from \cite{KGB})}
	\end{minipage}
\end{figure}

We may reformulate the situation as follows: the rotation curve of a galaxy displays the gravitational attraction generating the centripetal force that keeps the galaxy together. This amount of gravity does not correspond to the visible matter of the galaxy, being instead much larger. This discrepancy between gravity and visible mass is a "missing mass phenomenon" and constitutes an evidence of dark matter. In a greater or smaller amount, this phenomenon exists for almost all disc galaxies\footnote{\label{foot1} A notable exception is provided by the so-called ultra-diffuse galaxies (UDG) that can be as large as a typical disc galaxy but that contain only very few and very old stars. Indeed, some of these galaxies consist almost entirely of dark matter, while others appear to be almost entirely free of dark matter. We will discuss these features of the UDGs in Subsection 1.3.6.1.}. \\
From the rotation curves we can deduce not just the entire amount of dark matter in the galaxy, but its distribution as well. It turns out that in the central region the presence of dark matter is not necessary, because the rotation curve follows quite well the Newtonian prediction, corresponding to the visible matter. Instead, what is required is a \emph{dark matter halo}, surrounding the galaxy, that becomes relevant at the periphery and proceeds to large distances from the centre, where the luminous matter is very sparse. Indeed, recalling the analogy with the Solar System, we would expect that, for distances from the centre larger than the radius of the visible matter, most of the mass of the galaxy should essentially be taken into account (exactly as almost the whole mass of the Solar System is concentrated in the Sun) and that therefore the speed of revolution of stars and gas should start decreasing from there like the inverse of the square root of the distance (as it happens for the speeds of the planets, following Kepler's Third Law \cite{macdougal 2012}). However, this is not what we measure. In the outskirts of the galaxy and beyond, the rotation curve does not decrease at all, it is flat! In other words, instead of decreasing, the speed of revolution remains essentially constant as a function of the distance from the centre, apart from some small oscillations (see Figure \ref{1.1}). The Third Kepler Law, established inside the Solar System and being one of the greatest successes of Newtonian dynamics, is blatantly violated for much greater self-gravitating systems, such as disc galaxies. If there was no dark matter enveloping the galaxy, the observed non decreasing peripheral rotation velocity would imply the sparse luminous matter at the outskirts of the galaxy to quickly disperse into space. Instead this is precisely prevented by the presence of a dark matter halo, whose mass is very large, amounting to about $80$\% of the total mass of the system. It is certainly not a coincidence that all standard disc galaxies studied so far display strikingly similar in shape dark matter halos. \footnote{ For elliptical galaxies the dispersion of the velocities of stars renders the use of Keplerian motions impossible, and the evidence for dark matter comes instead from virial measurements, the measurement of the gas velocity in external regions, the effects of weak gravitational lensing, etc. From these observations it is not clear whether the dark matter forms a halo around the baryonic matter or whether it is more or less homogeneously mixed with the latter.}

The enigma is additionally complicated by a noteworthy discovery. To wit, call v$_f$ the rotation velocity in the flat region and let $M$ denote the luminous mass (e.g. the baryonic mass) of the galaxy. If one measures these data for all disc galaxies, one finds that $M$ is proportional to a certain power v$_f^{\alpha}$, with exponent $\alpha\cong3.5\div4$. This is the so-called Tully-Fisher relation. It is purely phenomenological, unexplained even in the dark matter paradigm. It would seem quite a remarkable coincidence that dark matter should accumulate around each galaxy in an amount sufficient to satisfy exactly such rule.

\subsubsection{Virial of clusters}\label{DM:2.2.1.2}

Not all galaxies have the form of a disc. Many are ellipsoidal or irregular in shape. For these galaxies one cannot reconstruct a rotation curve, since the velocities and the orbits of the stars are very complicated, and in general are not subject to some overall regular pattern. \\
The same happens in galaxy clusters and super-clusters where individual galaxies are also subject to complicated motions relative to one another. However, even for these more complicated systems it is possible to obtain a valuation of the amount of the gravitational mass of the system. The "Virial Theorem" connects the average speed of a gravitating body to its gravitational energy. The story is always the same: gravity keeps the system together, so that its gravitational mass can be estimated from the knowledge of the average internal velocities of its bodies. Then again, the mass so measured turns out to be too much compared to the luminous matter observed. As before, about $80$\% of the total mass of the system must be attributed to some unknown form of invisible matter, in order for things to work out correctly.

Historically, this has been one of the first motivations adduced to infer the existence of dark matter, when the Swiss astronomer Fritz Zwicky applied the Virial Theorem to the Coma Cluster in 1933 \cite{zwicky1, zwicky2}. He found that the cluster could hold together without disintegrating, with each individual galaxy flying off, only if it had a gravitational mass $400$ times bigger than the one which could be seen. Further studies take into account additional matter, like gas, but even the best present estimates tell us that the visible mass is five or six times less that the one found using the Virial Theorem applied to the average velocities of the cluster's galaxies.

\subsubsection{X-rays emission from clusters}\label{DM:2.2.1.3}

As stated earlier, galaxy clusters contains intergalactic gas and dust. This gas is extremely rarefied, ionized and hot. Like any hot material, it emits a radiation that is the more energetic the hotter is the gas\footnote{\label{foot2}It is known that, when sufficiently heated, scorching iron becomes red. Then, if heated further, it may acquire a light blue color corresponding to the emission of even higher energetic radiation. This mechanism works even at lower temperatures. For example, human bodies emit thermal radiation that is not sufficiently energetic to be seen by our eyes, but it can be perceived using infrared glasses.}. The intergalactic gas is so hot that it emits X-rays, a very energetic radiation. From the measurement of these X-rays, we can determine the gas' temperature. Now, the higher the temperature, the higher is the average velocity of the gas particles and the larger the velocity dispersion. Then, by the Virial Theorem these quantities are again related to the gravitational energy keeping the system together. If the mass were just the one we could observe, the inner velocities of the gas particles would be so high that the whole gas would quickly disperse, in spite of the gravitational attraction. Instead, since the system is stable, we can estimate the amount of mass that prevents it to disperse and we find a gravitational mass that is again five or six times larger that the one we observe, thus pointing again to the fact that dark matter accounts for about $80$\% of the total gravitational mass of the system.

\subsubsection{Gravitational lensing}\label{DM:2.2.1.4}

Recall that Einstein's Equations of general relativity imply that any distribution of mass and/or energy curves spacetime. This feature provides us with a further technique to measure mass. The large mass of a galaxy or of a galaxy cluster curves the space around it. This causes any light traveling close to the system to glide freely along the corresponding induced curved space and thereby getting deflected, as if the deflecting object were a lens. This happens because, even though light travels along straight lines, even the straightest possible line in a curved space is, by necessity, curved. In general relativity curved spacetime tell the objects, including light, how to move. For light the resulting effect is a so-called gravitational lens\footnote{\label{foot3}The bending of light passing close to a massive object was one the three crucial tests of general relativity that Einstein predicted when he first published his theory in 1915. The first of such tests was the explanation of the 43 seconds of arc per century anomalous (e.g. non-Newtonian) precession of the perihelion of the planet Mercury, the third one the frequency shift of light travelling in a gravitational potential. While the second one was precisely the bending of the light from stars, passing close to the Sun. This effect was verified in 1919 upon the observation of the light coming from stars close to the disc of the Sun during a total solar eclipse \cite{eddington 19}. Since those times innumerable more precise confirmations of general relativity in different contexts have taken place.},\footnote{\label{foot4}The first gravitational lens was discovered in 1979 when two close quasars were spotted in the sky having surprisingly very similar spectra \cite{gravlens}. It was then soon realized that one was dealing with two images of the same quasar, whose light had been split into two paths by the lensing effect of the gravitational action of an intervening galaxy.}. \\
If we are fortunate enough, it may happen that a background object as as a galaxy or a quasar happens to be right behind a foreground galaxy or a galaxy cluster that we are observing. Then we see the image of the background object enhanced, distorted, and often multiplied by the bending effect of the gravitational field of the foreground object. The larger is the mass of the foreground object, the greater are these distorting effects\footnote{\label{foot5}Occasionally, we can even be so lucky that the two objects, the background and the foreground one, are so perfectly aligned along our line of sight, that the lensed image of the background object is a circular ring (the so called Einstein Ring), whose radius is proportional to the mass of the lens.}. Now, we can in general accurately reconstruct the distribution and density of the mass of the lens by the shape of the image (or images) of the lensed object. In all such instances, it again invariably turns out that the lens is surrounded by a halo of invisible matter, whose mass is five or six times larger that the one of the lensing matter that we see.

\subsubsection{The Bullet Clusters}\label{DM:2.2.1.5}

The so-called Bullet Clusters are two known examples of pairs of galaxy clusters in which the two members of each pair have collided in the past. For reference, they have been designated respectively as 1E 0657-588 and MACS J0025.4-1225. They provide the best evidence to date for the existence of dark matter. Indeed in each of these two systems, the stars of the galaxies of the component clusters mostly passed right through during the collision, without being too much slowed down by their gravitational interaction. On the other hand, the hot gas of the two colliding components, seen in X-rays, interacted electromagnetically during the collision and therefore did slow down by attrition much more that did the stars. This intra-cluster gas provides most of the mass of the normal (baryonic) matter of the cluster pair. Therefore, if there was no dark matter in the system, one would expect the largest part of the mass of the latter to be found in the volume occupied by the gas. Now, in both the examples considered, the mass distribution can be reconstructed by the lensing effect produced by the colliding clusters on the background objects. And what one finds in this manner is instead that by far the greatest part of the mass is concentrated in the regions occupied by the clusters that have collided. Thus indicating that each of the colliding clusters has been carrying along his own dark matter halo. The two halos being presumably composed of only weakly and gravitationally interacting particles, having themselves passed right through essentially unaltered during the collision.

\subsection{Cosmological dark matter evidences}\label{DM:2.2.2}
We will now look at the dark matter phenomena that emerge at cosmological scales.  They concern the best model elaborated to date for the description of the evolution and the history of the universe as a whole. It is called the Cosmological Concordance Model, or $\Lambda$CDM (Lambda Cold Dark Matter) model. As we have seen in the previous section, this model is similar to the earlier Friedmann one, filling the universe with certain components that are almost uniformly distributed and that lead to the expansion of the universe according to general relativity. Differently from the original Friedmann model, $\Lambda$CDM accounts for a suitable amount of dark energy, alongside the traditional components of the universe, matter and radiation, in order to account for the observed accelerated expansion (see Section 2 of the Chapter on dark energy). This dark energy is usually assumed to act as a cosmological constant $\Lambda$, thus appearing in the acronym. The remaining letters, standing for ``Cold Dark Matter'', are there to indicate the necessity to add to the model a suitable amount of dark matter, in order to justify the evidences that we will expound in the following. What exactly it means that dark matter is assumed to be cold will be explained in Subsections \ref{DM:2.3.3} and \ref{DM:2.3.4}. 

\subsubsection{Fitting the deceleration parameter}\label{DM:2.2.2.1}
In Section 2 of the Chapter on dark energy we have seen how Perlmutter, Schmidt and Riess have been able to measure two important cosmological parameters, the rate of expansion $H_0$ and the acceleration parameter $q_0$, from the redshift of the light emitted by type $Ia$ supernovae. We have seen how the knowledge of $q_0$ allows one to estimate the value of the cosmological constant $\Lambda$, in case we wish to interpret the data with it. Anyway, not just $\Lambda$ but also another parameter of the $\Lambda$CDM model can be estimated from these measurements: the total amount of matter in the universe. As mentioned in Section 1 of the Chapter on dark energy, a Friedmannian model describes the expansion law of the universe by a function $a(t)$, once a few numbers have been assigned: the curvature of space, the average density of matter and radiation and, in the case of $\Lambda$CDM, also the density of an additional component of the universe, namely dark energy. That's why one says that the model has four parameters. Two of them, the curvature and the density of radiation, are determined to be essentially zero from other considerations. Hence, the other two measurements, of $H_0$ and $q_0$, fix the remaining parameters that are the percentages $\Omega_{M0}$ and $\Omega_{\Lambda0}$ of mass and dark energy respectively.\\
Since the other components are negligible, these two components sum up to provide the content of the entire universe, so that $\Omega_{M0}+\Omega_{\Lambda0}=100\%$. We have seen that the two effects compete: $\Omega_{M0}$ would decelerate the expansion, whereas $\Omega_{\Lambda0}$ would accelerate it. Then, according to Friedmann equations, the total deceleration today is $q_0=\frac 12 \Omega_{M0}-\Omega_{\Lambda0}$. $q_0$ has been measured to be $q_0\simeq -0.53$. From this it is a simple exercise of algebra to obtain the values $\Omega_{M0}\simeq 31.5\%$ and $\Omega_{\Lambda0}\simeq 68.5\%$.\\
Fitting the Friedmann equations we get an estimate of the total gravitational mass. As noted, this is the $31.5\%$ contribution $\Omega_{M0}$ to the total mass content of the universe, the remaining $68.5\%$ being accounted for by dark energy. As we did in subsection \ref{DM:2.2.1} 
for the astrophysical evidences for dark matter, we can check here too if this overall contribution to gravity matches with the amount of matter that we can see (the visible matter), or if there is again some ``missing mass''. Taking into account all stars, any other type of astrophysical bodies (black holes, neutron stars, planets etc.), dust and gas, we arrive at establishing that the average density of normal matter is about $0.2\sim0.25$ atoms per cubic metre. This is a very small amount indeed. It is only about $4.4\%$ of the total mass-energy content of the universe\footnote{\label{foot6}The contribution of radiation to this amount is negligible. But what about neutrinos? The present total density of all neutrinos is about $0.336\times10^9$ per cubic metre. The known upper limit of neutrino mass is of the order of one millionth the mass of the electron (to be precise, $0.8$ electronvolt). This means that the contribution of neutrinos to the gravitational mass could be up to $1/5$ the contribution by atoms, not a negligible amount. Of course, this figure is an upper limit, neutrino masses could be much smaller.}. Comparing with the value $\Omega_{M0}=31.5\%$, it turns out that $27.1\%$ of the remaining amount should be dark matter. In other words, denoting by $\Omega_{DM0}$ the percentage of dark matter, we get $\Omega_{DM0}/\Omega_{M0}\simeq 86\%$, a figure that is coherent with the dark matter abundances in galaxies and clusters.\\
To summarise. It is remarkable how big the Dark Sector is! All our measurements and observations indicate that, out of the whole amount of matter and energy in the universe only about $4.4\%$ of the total is made of the normal (baryonic) matter that we are familiar with: the stuff of stars, planets and our own bodies. The remaining $95.6\%$ is made of dark energy and dark matter, two entities to date still draped in mystery.

\subsubsection{Formation of structures}\label{DM:2.2.2.2}
As we have stated, the $\Lambda$CDM model is built on the hypothesis of the universe's homogeneity at sufficiently large scales, say roughly of the order of one hundred million light years, or so. However, the model should be able to justify also the progressive emergence of inhomogeneities at smaller scales. In other words, where do stars, galaxies and other aggregate of matter come from? The Cosmological Model does indeed describe how these structures gradually grew thanks to gravitational attraction. To wit, in the primordial universe there were present small inhomogeneities such as regions where matter was slightly more dense than in their immediate surroundings. 
By the action of their stronger local gravity, these regions attracted more mass from their neighbouring regions, thus becoming even more dense. This, in turn, attracted more additional mass, increasing their density further. Whereas, by contrast, in the lower density regions, the expansion of space was less slowed down by the weaker gravity, so that these regions expanded further, their density decreasing even more. Ultimately, these self-catalytic mechanisms have lead, through the competing effects of matter and dark energy, to the universe in its present state: at scales larger than those of galactic clusters, it is structured in huge ``void bubbles'' bordered by ``filaments'' and ``walls'' made by clusters and superclusters.\footnote{\label{footnotona}Here, the reader may ask: where did the original small inhomogeneities arise in the first place? Cosmologists agree on the following picture. In the very early universe, close to the Big Bang, at the scale of the Planck length $l_p=\sqrt{\frac {hG}{2\pi c^3}}$ (see footnote 28 of the Chapter on dark energy), 
quantum effects were important and it was the quantum fluctuations that gave rise to the original inhomogeneities. This was followed by the extremely short \emph{inflationary era} \cite{brandemberger}, during which the universe experienced a tremendous exponential growth which, among other things, almost flattened space and eventually subsided, being replaced by the normal expansion. By virtue of the huge exponential stretching provoked by inflation, the original tiny quantum inhomogeneities grew to macroscopic sizes. Then, dark matter and ordinary matter, the latter after matter-antimatter annihilation and the formation of baryons (footnote 16 of the Chapter on dark energy) from the residual matter (baryogenesis), started to accumulate inside the denser areas, creating overdense local regions. It is from these clumpy regions of matter that baryon sound wave oscillations (the BAO, see subsection 2.1 of the Chapter on dark energy) emerged radially, propagating inside the primordial plasma and eventually freezing in place after the decoupling of matter and radiation, about $380,000$ years after the Big Bang. Whereas dark matter, which did not interact electromagnetically, stayed put in the overdense regions (see Fig. 1 of the Chapter on dark energy). These configurations are still printed and visible in the CMB. And it is this intricated network of frozen BAO and overdense dark matter clumps that triggered the later formation of structures (stars, galaxies, galaxy clusters and superclusters).}
Of course, these collapses must have been sufficiently efficient to indeed give rise to the hierarchical clustering that we see today, during the $13.8$ billion years that have elapsed since the Big Bang. However, if one performs computer simulations with the matter that we can see, one finds that the latter is not enough to allow for the formation of the present structures in such a time lapse. Instead, from this to happen, one needs again a substantial amount of missing mass: dark matter. If one adds to the simulations the right contribution of dark matter, one finds that the resulting gravity guarantees sufficiently fast collapses. Indeed, originating from the network of overdense dark matter regions and the frozen BAO of baryonic matter (see footnote \ref{footnotona}), dark matter anisotropies started to collapse further onto themselves, generating dark matter halos. In turn, these attracted normal matter, which fell swiftly enough into the halos' gravitational wells. Once there, friction drove baryonic matter to further coalesce, giving rise to the first stars around $300$ million years after the Big Bang. According to this theoretical model, when the universe was about one billion years old stars grouped into the first galaxies. \\
However, we must add here that this scenario has been recently severely questioned by the findings of the James Webb Space Telescope (JWST), that was launched on December 25, 2021.\footnote{\label{foot8}The James Webb Space Telescope (JWST) is a $6.5m$ infrared telescope operating in the wavelength band $0.6-28.3$ micrometres ($\mu m$) (orange to mid-infrared). It was launched on 25 December 2021 and is on a halo orbit around the $L_2$ Lagrange point of the Sun-Earth system, at a distance of about $1.5$ million Km from Earth. It is mostly designed for observations of the first stars and the first galaxies, and the study of the atmospheres of potentially habitable exoplanets.} Indeed, among its already innumerable stunning images and findings, whose illustration is out of the scope of the present article, JWST has spotted several disc galaxies as large as the Milky Way or more \cite{coley}, that were already present at such early dates as around $400$ million years after the Big Bang. This, among others, such as the Hubble tension (see footnote 12 and Subsection 3.4 of the Chapter on dark energy), the $\sigma-8$ tension \cite{abdalla}, the discovery of El Gordo galaxy cluster \cite{el gordo} (and other phenomena), are some of the findings that are challenging the validity of the $\Lambda$CDM model. Though, at least as regards the early formation of such large and regular galaxies, the phenomenon may possibly be explained by the fact that at such early times the temperature and pressure of the baryonic matter fallen into the already formed dark matter halos were high enough to allow for a very fast formation of first generation stars and their subsequent quick grouping into large galaxies.

\subsubsection{The multipole spectrum of the CMB}\label{DM:2.2.2.3}
\begin{figure}
\label{Figure2}
	\begin{minipage}[b]{1.0\linewidth}
		\centering
		\includegraphics[width=7 cm]{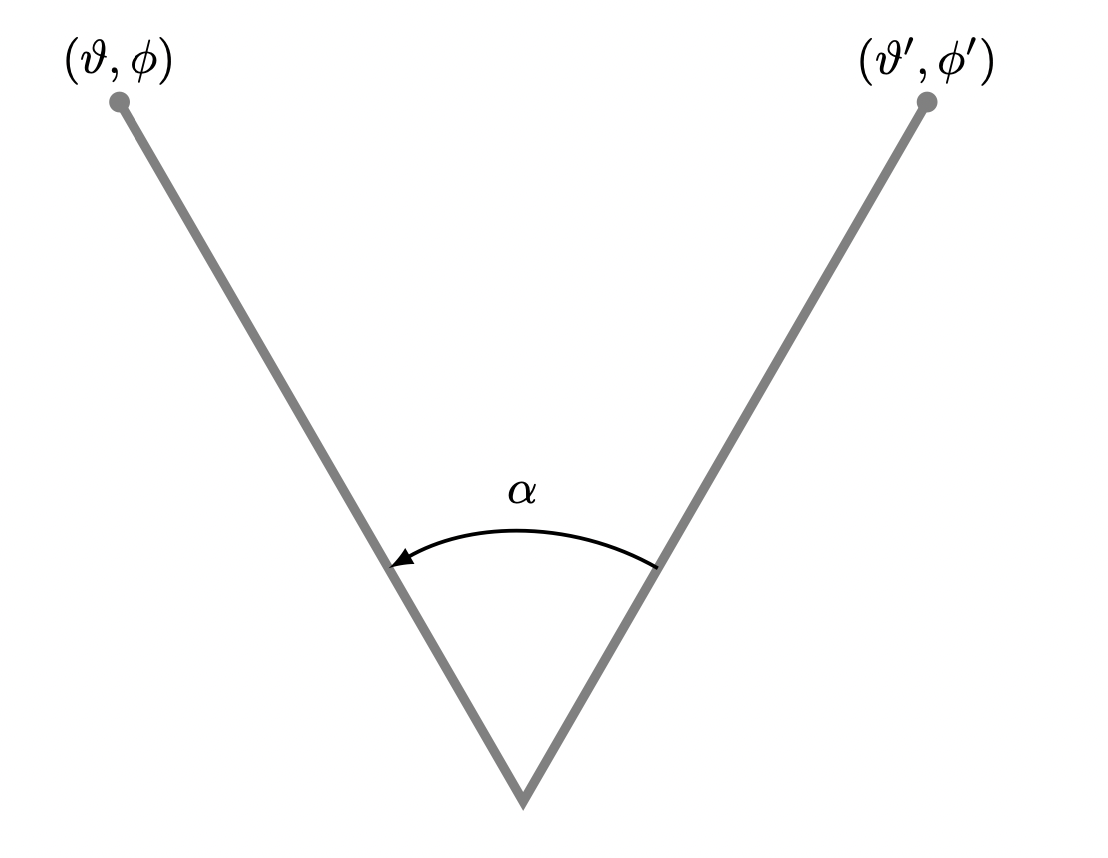}
		\caption{\label{angle} }
	\end{minipage}
\end{figure}

The Cosmic Microwave Background Radiation (CMB) (see subsections 1.3 and 2.1 of the Chapter on dark energy) is the thermal electromagnetic radiation originated at the time, about $380,000$ years after the Big Bang, when the temperature of the expanding universe had become low enough, around $3,000$K\footnote{\label{foot9}The Kelvin (K) scale is the temperature scale used in the world of science. Zero Kelvin (0K) is the lowest possible temperature (the so called absolute zero), corresponding to complete absence of thermal energy. The Kelvin scale uses the same unit increment of the Celsius scale, meaning that the temperature interval of one Kelvin (one K) is the same as the temperature interval of one Celsius (one ${}^\circ$C). The absolute zero (zero K) corresponds to $-273.15^\circ$C. Therefore, the temperature of the CMB, $2.73$K, corresponds to less than 270 degrees Celsius below zero.\\
The symbol $m$K (milliKelvin) denotes one thousandth of a Kelvin or $10^{-3}$K, the symbol $\mu$K (microKelvin) one millionth of a Kelvin, or $10^{-6}$K. }, for photons not to be able to ionise neutral atoms anymore. Photons did not scatter any more off baryons and electrons, hence matter and radiation were decoupled and the latter started to travel freely through space: the universe became transparent. Traveling to us from the so called \emph{surface of last scattering} (at $z=$1,100), the CMB has undergone an enormous redshift by the intervening expansion of the universe up to the present era, and its temperature now measures $2.73$K. However, the temperature of the CMB is not exactly uniform. It exhibits small variations of the order of a few tenth of micro K all over the sky. These variations are the imprint on the CMB of the original inhomogeneities in the density of the early universe (see \ref{DM:2.2.2.2} and footnote \ref{foot9}), those that formed the seeds for the later formation of the present day structures (galaxies, galaxy clusters and superclusters). Indeed, the regions of the early plasma that were slightly denser than the neighbouring ones, where also slightly hotter, and these small temperature differences were carried by the radiation as well, it being in thermal equilibrium with the plasma. The temperature anisotropies of the CMB were first revealed by the Cosmic Background Explorer (COBE) satellite \cite{smoot, smoot1} and later characterised on much smaller scales by the Wilkinson Microwave Anisotropy Probe (WMAP) \cite{komatsu} and more recently by the ESA Planck Surveyor Satellite \cite{planck}.
At any given point in the sky, characterised by its angular coordinates $\theta$ and $\phi$ ($0\leq\theta\leq\pi$, $0\leq\phi\leq2\pi$), the temperature fluctuations are conventionally measured by the difference between the value $T(\theta,\phi)$ of the temperature of the CMB at the point, minus the average temperature $\langle T\rangle$, and divided by $\langle T\rangle$:
$$
\frac {\delta T}{T}(\theta,\phi)=\frac {T(\theta,\phi)-\langle T\rangle}{\langle T\rangle}.
$$
Now consider two points of the celestial sphere (the surface of last scattering) of coordinates, say, respectively, $(\theta,\phi)$ and $(\theta',\phi')$, whose directions form a given angle $\alpha$: see fig. \ref{angle}.

Denote by $C(\alpha)$ the product $\frac {\delta T}{T}(\theta,\phi)\frac {\delta T}{T}(\theta',\phi')$ averaged over all points separated by the given angle $\alpha$:
$$
C(\alpha)= \left\langle \frac {\delta T}{T}(\theta,\phi)\frac {\delta T}{T}(\theta',\phi') \right\rangle_\alpha.
$$
The function $C(\alpha)$ ($0\leq\alpha\leq\pi$) provides a statistical measure of the correlation of the temperature fluctuations of the CMB for different values of the angle $\alpha$. To study how these temperature fluctuations are correlated as functions of $\alpha$ it is useful to expand $C(\alpha)$ in a series of the so called Legendre polynomials \cite{legendre} $P_l(x)$; $l=0,1,2,3,4,5,\ldots$; $-1\leq x\leq 1$, expressed as functions of the cosine of the angle $\alpha$:
\begin{align}
    C(\alpha)=\frac 12 \sum_{l=0}^\infty C_l P_l(\cos\alpha). \label{(1)}
\end{align}
The first few $P_\alpha(x)$ are:
\begin{align*}
    P_0(x)&=1, \qquad P_1(x)=x, \qquad P_2(x)\frac 12(3x^2-1),\\
    P_3(x)&=\frac 12 (5x^3-3x),\\
    P_4(x)&=\frac 18 (35x^4-30x^2+3),\\
    P_5(x)&=\frac 18 (63x^5-70x^3+15x),\ldots.
\end{align*}
Formula \eqref{(1)} is called the \emph{multipole expansion} of $C(\alpha)$. The coefficients $C_l$ are uniquely determined by $C(\alpha)$ and are called the \emph{multipoles} of the expansion: $C_0$ is the monopole, $C_1$ the dipole, $C_2$ the quadrupole, $C_3$ the octupole, \ldots, $C_n$ the $2^n$-pole, \ldots.
From the particular structure of the Legendre polynomials it turns out that, generally speaking, a term $C_l$ is a measure of $\delta T/T$ on the angular scale $180^\circ/l$. In loose words, this means that if we are interested in how temperature fluctuations are correlated for a given angle $\alpha$ between two directions on the surface of least scattering, this will be provided by the multipole $C_l$ for which $l\simeq 180^\circ/\alpha$.\\
The \emph{power spectrum} of the temperature fluctuations of the CMB is plotted as a function of $l$ (or, equivalently, as a function of $\alpha\simeq 180^\circ/l$) as
\begin{align}
    \Delta^2_T(l)=\langle T\rangle^2C_l. \label{(2)}
\end{align}
It is measured in microKelvin square ($\mu$ K$^2$). Multiplying \eqref{(1)} by the square of the average temperature gives
\begin{align}
    \langle T\rangle^2 C(\alpha)=\frac 12 \langle T\rangle^2 C_0 +\frac 12 \langle T\rangle^2 C_1 \cos\alpha+\frac 12 \sum_{l=2}^\infty \Delta^2_T(l) P_l(\cos\alpha). \label{(3)}
\end{align}
In \eqref{(3)}, the first term, the monopole, does not depend on $\alpha$ and is simply proportional to the square of the average temperature $\langle T \rangle=2.73$K. The second term, the dipole one, behaving like $\cos\alpha$, is the one that signals the motion of the Solar system inside the universal reference frame of the CMB: relative to the Solar system, and after correcting for the periodic motion of the Earth around the Sun, the CMB temperature is slightly higher towards the direction along which the Sun is moving ($\alpha=0$, $\cos\alpha=1$) and slightly lower in the opposite direction ($\alpha=\pi$, $\cos\alpha=-1$). The dipole temperature contribution $T(\alpha)$ to $\langle T\rangle$ is of the order of $\langle (\delta T/T)^2\rangle^{\frac 12}\simeq 10^{-3}$, just a few $m$K. We have thus 
$$
T(\alpha)\simeq \langle T\rangle \left( 1+\frac {\rm v}c \cos\alpha \right),
$$
where v$\simeq 370Km/s$ is the speed of the Solar system within the CMB, being the composition of the orbital speed of the Sun around the centre of the Milky Way and the speed of the proper motion of the Milky Way.

\begin{figure}
\label{Figure3}
	\begin{minipage}[b]{1.0\linewidth}
		\centering
		\includegraphics[width=8 cm]{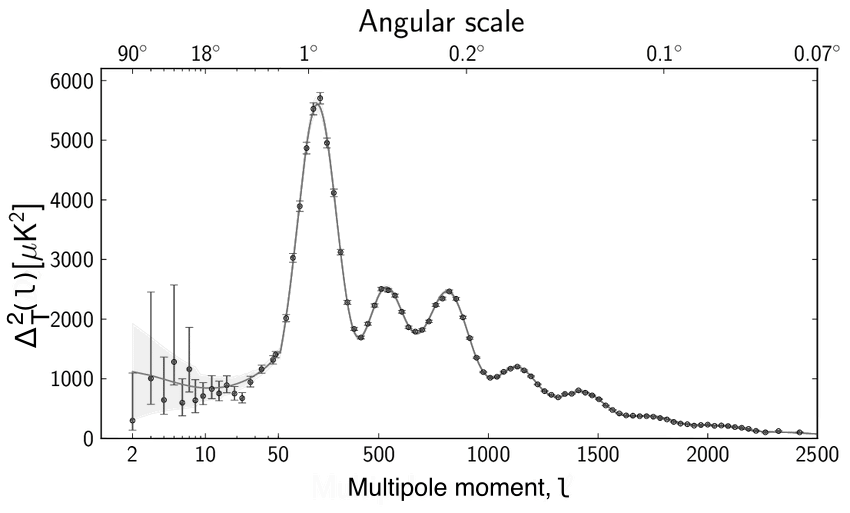}
		\caption{\label{2.3} Graph of the power spectrum \eqref{(2)} of the temperature fluctuations of the CMB, measured by the Planck satellite. The large error bars at the left part of the graph reflect the fact that, at large angular scales (large values of $\alpha$), there are fewer pairs of directions on the celestial sphere, as depicted in Fig. \ref{angle}. Note that the characteristic values in $\mu$K$^2$ appearing in the $y$-axis of the graph reflect the fact that the temperature anisotropies of the CMB are typically of the order of a few tens of a millionth of a Kelvin. (Adapted from Plank Collaboration Data, https://sci.esa.int/web/planck/-/51555-planck-power-spectrum-of-temperature-fluctuations-in-the-cosmic-microwave-background)}
	\end{minipage}
\end{figure}

The power spectrum of the temperature fluctuations that allow us to determine the values of the various cosmological parameters is thus encoded in formula \eqref{(3)} for $l\geq2$. After corrections are made for various effects that photons undergo during their travel to us on Earth from the surface of last scattering (such as gravitational lensing effects, the Sunyaev-Zel'dovich effect \cite{sunyaev, sunyaev1} and several others) one gets the graph of Fig. \ref{2.3},
where the values of $l$ and $\alpha$ are on logarithmic scale. Without going into the (though important) details, we will confine ourselves to the discussion of the three mean peaks of the graph, at the values $l\simeq 200$ ($\alpha\simeq 0.9^\circ$), $l\simeq 520$ ($\alpha\simeq 0.33^\circ$), $l\simeq 900$ ($\alpha\approx 0.2^\circ$) respectively.

To interpret the meaning of these peaks, we need some preliminary notions. Shortly before decoupling, the composition of the universe consisted of a mixture of photons, protons, electrons, helium (He) nuclei, neutrinos and dark matter. Practically, neutrinos and dark matter did not interact with the rest. Contrary to our present epoch, at which the energy density of the photons is negligible compared to the energy density of baryons, at that time it was dominant. Photons and baryons were forming a plasma in which the pressure exerted by the photons was competing with the gravitational attraction causing the fluid to fall into the potential wells of the overdense regions of dark matter clumps. As the fluid fell into the wells, it got compressed, until the increased pressure would push the fluid back up, having it expanding. The expansion lowered the pressure to the point where the fluid would again start falling back into the well and getting compressed, with the whole cycle starting all over again. These alternations of compression and depression, created sound waves that can be shown to propagate in the fluid at the speed v$=c/\sqrt 3$. They are precisely the Baryon Acoustic Oscillations (BAO), discussed in Subsection 1.2.1 of the Chapter on dark energy. At the time of decoupling, the photons where released, hence the BAO froze in place and they remained imprinted in the CMB. As discussed in subsection 1.2.1 of the Chapter on dark energy, each BAO propagated outward, from some original overdense region, up to a spherical surface of compression (a crest) on which it froze at decoupling (see Fig. {1.1} of the Chapter on dark energy). Any such surface is called a \emph{sound horizon}. Now, each sound horizon on the surface of last scattering gives rise exactly to the \emph{first peak} of Fig. \ref{2.3}. Combining the radius of any sound horizon (the ``standard ruler'') and the angular scale ($\alpha\simeq 0.9^\circ$) of the first peak allows one to conclude that the space curvature of the universe is practically flat, giving credit to the inflationary scenario (see footnote \ref{footnotona}). Indeed, if the space were positively, or negatively, curved, the position of the first peak in Fig. \ref{2.3} would be located more to the left or, respectively, more to the right than its actual location. \\
The \emph{second peak} corresponds the spherical BAO wave of depression (a trough), and as such, it signals the various concentrations of baryons on the verge of falling back into the dark matter wells.\\
As to the \emph{third peak}, its amplitude and shape provides information on the overdense regions and on the amount of dark matter they contain, and from which the BAO originated. In particular, this information gives another, independent indication that the amount of dark matter in the universe is between 5 and 6 times larger than the amount of normal matter, in agreement with the astrophysical and cosmological evidences that we discussed earlier.\footnote{\label{foot10}Actually, one would expect further peaks to appear to the right of the third peak in the graph of Fig. \ref{2.3}, corresponding to higher harmonics of compression and depression. However, one sees that this is not the case: one observes just two bumps, and then the correlations of the temperature fluctuations drop off fast at increasingly smaller scales. The main reason for this is that at smaller scales the mean free path of photons becomes relevant, hence the latter diffuse and quickly smooth out the temperature anisotropies.}

\section{Hypotheses on the missing mass}\label{sec:missingmass}\label{DM:2.3}
In the last section we have discussed the observational facts that suggest the existence of dark matter. In the present section we will deal with the leading hypotheses that have been put forward for the interpretation of this remarkable and mysterious phenomenon. We speak about hypotheses because no satisfactory solution to the problem has been found to date. These hypotheses suggest alternative explanations that are mostly mutually incompatible. Actually, some of them even deny the existence of dark matter and, instead, hypothesise a more complex dynamics. This is the reason why, in general, it would be better to speak, in the sequel, of \emph{missing mass}, in place of dark matter. The hypothesis that this missing mass does actually consist of some form of invisible matter is at present the preferred one by the majority of the scientific community. Hence, it will be the first one that we will talk about. However, in the critical attitude that should always permeate scientific research, it is crucial to discuss the other hypotheses as well, as long as one, more, or even all of them may be disproved by observations (including of course the leading one, should this turn out to be the case).\\
Before we describe the hypotheses at the basis of dark matter, we recall the constituents of the visible matter, that are well described by the Standard Model of Elementary Particles. In the sequel, we will refer to the latter simply as the Standard Model.

\subsection{The Standard Model}\label{DM:2.3.1}
The Standard Model describes the different known classes of elementary particles and their interactions \cite{standard model}. These are the simplest objects that physics has been able to identify, and they mutually combine to form any other existing thing. The situation is similar to the one set up in chemistry. Atoms are the elementary components of chemical reactions, but they are not the most elementary objects for what concerns physical phenomena. Indeed, atoms are composed by electrons, proton and neutrons, kept bound together by the electromagnetic and nuclear forces. Protons and neutrons are not themselves elementary: they are in turn composed by combinations of quarks, kept bound together by the so called colour interactions, or (quantum) chromodynamics, the basis of all nuclear forces. The structure of matter does not break down into further elementary components: to the best of our knowledge, quarks and electrons are elementary particles. They are not the only ones, however. During the 20th century other elementary particles have been found: neutrinos, other types of quarks; and a kind of ``cousins'' of the electron, though heavier and unstable, the muon and the tauon.\footnote{\label{foot11}Neutrinos do not have electric charge and exist in three types, associated to the electron and its heavier counterparts: the electron neutrino, the muon neutrino, and the tau neutrino. Since all these six particles (with the exception of the tauon) are lighter than the proton and the neutron, they are collectively called \emph{leptons}, from the Greek $\lambda\varepsilon\pi\tau{\acute o}\varsigma$ (scarse, light, small). Each of these particles has its corresponding antiparticle (see Subsection \ref{DM:2.1.1}). All leptons are fermions (see footnote 30 of the Chapter on dark energy), and their spin is $1/2$. Also each type of quark has of course its corresponding antiparticle.} And finally, at the beginning of our century, the celebrated Higgs boson.\\
These particles interact among themselves through four fundamental forces. In order of increasing intensity: gravity, the weak nuclear force, electromagnetism, and the colour force (chromodynamics, or the strong nuclear force). Aside from gravity, these particle and forces are described by interacting quantum fields (see subsection 3.2 of the Chapter on dark energy). Each force of the Standard Model is transmitted by a certain particle. For example, the electromagnetic force is transmitted by the photon, the excitation (quantum) of the electromagnetic field (see subsection 3.2 of the Chapter on dark energy). The photon is a ``vector boson'', since its spin is 1 (see footnote 26 of the Chapter on dark energy; bosons having spin zero are called ``scalar bosons''). The transmitters of the colour force and of the weak nuclear force are vector bosons as well. The particles transmitting the colour force are called gluons and there are eight types of them. Whereas there are three types of transmitters of the weak nuclear force, the so called intermediate vector bosons, denoted by $W^+$, $W^-$ and $Z^0$.\\
Quarks, like leptons, are spin $1/2$ fermions. They interact among themselves so intensely, by exchanging gluons, that it is impossibly to separate them. Therefore, they exist in particles composed by quark-antiquark pairs (bosons called mesons) or as triplets of quarks (fermions called baryons, protons and neutrons are triplets of this kind), cumulatively called hadrons. The strong interactions is an exclusive prerogative of quarks, whereas both quarks and leptons (as well as the Higgs boson) can interact both weakly and electromagnetically, with the exception of neutrinos that, having no electric charge, interact only weakly.\footnote{\label{foot12}Actually, though neutral, it is not excluded that neutrinos may have a magnetic moment, though extremely small. Indeed, it is estimated that, if they have one, it would be of the order of $10^{-19}\sim 10^{-20}$ times the magnetic moment of the electron. Such being the case, the electromagnetic interaction of neutrinos, if any, would to all purposes be nonexistent.} In addition, it is known that the three types of neutrinos have different masses, but it is not known if they have all a mass. At any rate, their mass is so small that, to date, only an upper limit of their mass is known: $0.8eV$ (for the definition of $eV$ see footnote 29 of the Chapter on dark energy). It is important to note that almost all particles are unstable (including the neutron, and perhaps the proton, though so far no proton has yet been seen to decay), and they decay into lighter particles. However, neutrons and protons appear to be stable inside non radioactive atoms. This is the reason why, neutrons and protons being collectively called baryons (see footnote 16 of the Chapter on dark energy), the visible matter is called baryonic.

\subsection{Hot dark matter}\label{DM:2.3.2}
Among other things, dark matter is supposed to have had important role already since the primordial epochs after the Big Bang, for example in the formation of anisotropies or structures. Now, all known particles have masses never larger than a few hundred $GeV$ (see footnote 29 of the Chapter on dark energy) in energy units (recall that the energy content of a particle of mass $m$ equals $mc^2$). Or, since special relativity implies that a particle moving with speed v has energy equal to $mc^2/\sqrt{1-{\rm v}^2/c^2}$, it follows that, whenever a particle is moving with a speed very close to the speed of light (as for instance in particle accelerators or in energetic cosmic rays) its rest energy $mc^2$ may become negligible compared to its total energy. Then, were dark matter made of some kind of particles, it would make no exception. Thus, the point is whether hypothetical particles of dark matter have a sizable rest mass or a very small one like neutrinos, and whether they move slowly or at relativistic speeds. Consider first the hypothesis that dark matter particles have a small mass. Then, the fact that the missing mass of the universe largely exceeds the baryonic mass, would imply that these particles are either present in an utterly unimaginable number or, much more likely, are ultrarelativistic, namely they form a very hot gas. In this case, we would speak of \emph{hot dark matter}. If this were the actual situation, neutrinos would apparently be an optimal candidate for hot dark matter, since they only interact weakly and gravitationally. However, since their mass is small, one would need an enormous number of neutrinos to justify the total amount of DM in the universe. Now, the Standard Model allows us to calculate the number of neutrinos existing in the visible universe, starting from an estimation of the visible matter, thanks to the fact that it describes in a very precise way the interactions among neutrinos, the other fermions and the Higgs boson, hence the mechanisms of generation of the former. However, the number so obtained, multiplied by the estimated upper limit for the neutrino mass, leads to a much lower figure, viz. to only a fraction of the visible baryonic mass.\\
A possible idea, suggested by the so called grand unification theories, introduced to solve some of the imperfections of the Standard Model, is that there could exist another type of neutrinos, called \emph{sterile neutrinos} \cite{sterile neutrinos}. Ordinary neutrinos are left handed, namely they are always produced with their spins antiparallel to their direction of motion. On the other hand, sterile neutrinos, if they exist, would be right-handed, meaning that, instead, their spin would always be parallel to the direction of their motion. As such, sterile neutrinos would not interact with baryonic matter (hence the adjective sterile), and their action would be only gravitational. In particular, their quantity would be uncorrelated to the amount of visible matter. However, in most cases dark matter seems to be more concentrated in halos around galaxies, more that around galaxy clusters and superclusters, hence too much localised to correspond to hot dark matter. Indeed, galaxies are too small objects to be able to capture gravitationally a gas of very light particles, that move practically at the speed of light. Therefore, at least concerning phenomena of galactic nature, we should expect dark matter instead to be made up of sufficiently slow (and sufficiently massive) particles, thus having to do with cold dark matter. On the other hand, we have seen that astrophysical and cosmological phenomena both substantially foresee the same amount of dark matter. This suggests the idea that such matter should mainly be of the same nature on both scales, and hence that it be mostly cold. Indeed, cold dark matter would act on both scales, whereas hot one would only act on cosmological ones, hence, if it where not present in negligible amount compared to the former, it would imply greater effect on cosmological scales compared to the galactic ones.

\subsection{Cold dark matter: MACHOs}\label{DM:2.3.3}
We refer the term cold dark matter to invisible bodies moving with subrelativistic velocities. If these are particles, they should have a sufficiently sizable mass and, in order not to be seen, they should interact only weakly and gravitationally, hence, in particular, they should be neutral. Hypothetical particles of this kind have been termed WIMPs, acronym for Weakly Interacting Massive Particles. We discuss WIMPs in the next Subsection. Instead, in the present one, we will consider the contribution to dark matter coming from non luminous macroscopic bodies, for which the term that has been coined is MACHOs, acronym for Massive Compact Halo Objects. For sure, galaxies are populated by a host of objects of this kind: isolated black holes, isolated neutron stars, white dwarfs, brown dwarfs, rogue planets, and possibly smaller objects like wandering asteroids. Rogue planets are planets that do not orbit a star, hence they move freely in space, far away from any luminous source. Most likely, they formed in protoplanetary discs of gas and dust surrounding newly born stars, during the process of formation of a planetary system, and have been expelled from the disc by some slingshot effect arisen from their interaction with other newly formed planets in the disc. They may be rocky (the smaller ones) or gas giants, their mass varying from the mass of Mercury or less, up to 13 times the mass of Jupiter.\footnote{\label{foot13}Jupiter is the largest planet in the Solar System. Its mass amounts to 318 times the mass of the Earth. It is what is called a gas giant: it has a small rocky core, about 20 times the Earth's mass, while the rest is almost entirely composed of hydrogen and helium.} In gas giants, the temperature and pressure in the core, though high, are not enough to trigger the fusion of hydrogen or of heavier elements, hence the body doesn't burn any fuel. By contrast, brown dwarfs are substellar objects whose range of masses runs from 13 to about 80 Jupiter masses. They are still not massive enough for hydrogen in their core to fuse into helium and heavier elements, making them into stars. However, their temperature and pressure in their centre is high enough to trigger the fusion of deuterium.\footnote{\label{foot14}In nature, a given atom is characterised by the number of protons in its nucleus, denoted by $Z$ and called \emph{atomic number}. Unless it is ionised, an atom is neutral, hence the number of electrons orbiting the nucleus is equal to the number of protons. However, a given atom can have different \emph{isotopes}, which differ one another by the number of neutrons in the nucleus. For example, carbon, whose symbol is $C$ and whose atomic number is 6, has three isotopes, ${}^{12}_{\ 6}C$, ${}^{13}_{\ 6}C$ and ${}^{14}_{\ 6}C$, having respectively 6, 7 and 8 neutrons in the nucleus (the number at the upper left of the symbol, denoted by $A$, is the total number of \emph{nucleons}, namely the number of protons plus the number of neutrons; e.g. in ${}^{14}_{\ 6}C$, 6 protons $+$ 8 neutrons = 14 nucleons). ${}^{12}_{\ 6}C$ and ${}^{13}_{\ 6}C$ are stable isotopes, whereas ${}^{14}_{\ 6}C$ is unstable; it decays into nitrogen-14 plus an electron and an electron antineutrino, ${}^{14}_{\ 6}C \rightarrow {}^{14}_{\ 7}N+e^-+\bar\nu_e$, with a half life of about $5,730$ years. All isotopes of a given atom (element) have the same chemical properties. Hydrogen, denoted by the symbol $H$, is the simplest of all atoms, having a nucleus made of just one proton, orbited by one electron. Deuterium, denoted by $D$, is the hydrogen isotope whose nucleus is formed by one proton and one neutron.} Now, the ratio of deuterium to hydrogen in gas giants is 26 deuterium atoms per million hydrogen atoms. This means that the energy production by nuclear fusion in brown dwarfs is so low that these objects are practically invisible. They are aborted stars.\\
When, in interstellar giant clouds of gas and dust in a galaxy, a local region of higher density collapses gravitationally into a clump of matter heavier than about 80 Jupiter masses, the temperature and the pressure in the core of the clump becomes high enough to trigger the fusion of hydrogen into helium and a star is born. Thus, in order to have a comparison with the mass of our star, since the mass of Jupiter amounts to $0.09\%$ the mass of the Sun, we realise that the lightest stars have masses of about $7.2\%$ the mass of the Sun. On the other hand, the most massive stars have masses that can reach values up to $200\sim250$ solar masses. Thus we see that the range of star masses is very wide, the heaviest stars having masses well more than three orders of magnitude larger than the lightest ones.\\
For intermediate mass stars like the Sun, or having masses up to $8\sim10$ solar masses, when the reservoir of hydrogen in their cores is exhausted, these stars die through processes briefly described in section 2 of the Chapter on dark energy, eventually leaving behind hot, collapsed, compact remnants, called \emph{white dwarfs}. These stellar cores, mostly composed by carbon and oxygen, are supported from further collapse by the pressure generated by their electrons, of quantum mechanical origin, called \emph{degeneracy pressure}.\footnote{\label{foot15}Electrons are fermions and, as a consequence of quantum mechanics, all fermions obey the so called Pauli exclusion principle, after the Austrian physicist Wolfgang Pauli: no two fermions of the same type can occupy the same (quantum mechanical) energy level \cite{massimi}. Hence, if one has a huge amount of electrons, say, in a body such as a white dwarf, even though the latter may be in a state close to its state of lowest energy, the electrons will pile up in states of increasing energies, that can thus become extremely large and therefore may generate an intense pressure. It is the latter that is called degeneracy pressure.} White dwarfs typically have masses between $0.17$ and $1.33$ solar masses, radii comparable to the Earth's radius, and densities of the order of one ton per cubic centimetre. When newly formed, their initial surface temperature is of the order of $100,000$ degrees Celsius. However, since in them no fusion takes place any more, they slowly radiate their heat into space, eventually becoming invisible cold remnants called \emph{black dwarfs}. As such, these objects can well give their contribution to the composition of dark matter.\\
Stars whose masses are greater than $8\sim10$ solar masses fuse heavier elements, up to iron.\footnote{\label{foot16}In any star it is in general impossible to fuse elements with atomic numbers larger than that of iron (the atomic number of iron is 27). Indeed, among all the elements of the periodic table, iron is the one that has the largest binding energy per nucleon. This means that any reaction that will fuse iron into elements of higher atomic number would be endothermic in place of exothermic. In other words, in order to trigger such a reaction one would have to furnish energy from outside. Hence elements heavier than iron, up to uranium, are never produced in stars by fusion. They are only produced in catastrophic events like the explosion of a supernova or a merging of two neutron stars or of a neutron star and a black hole.} They die catastrophically as supernovae, crushed by gravity no more balanced by the pressure generated by the energy produced by nuclear fusion in the core: gigantic explosions that scatter a lot of debris into space, flying off at speeds that can approach one tenth of the speed of light, and leaving as a central residual core either a neutron star or a black hole.\\
A neutron star can form with a mass not lower than about $0.8$ solar masses, but not exceeding much more than two solar masses. Indeed, as established theoretically by the Indian physicist Subrahmanyan Chandrasekhar, no stable white dwarf can exist with a mass larger than 1.4 solar masses (the Chandrasekhar limit \cite{chandrasekhar mass}). The reason is that, for masses larger than this limit the electron degeneracy pressure is not anymore able to balance the pull of gravity, and the body crushes further, as it happens for example with type $Ia$ supernovae (see Section 2 of the Chapter on dark energy). Neutron stars have densities of the order of the density of atomic nuclei, namely around one billion tons per cubic centimetre. In them, almost all of the electrons are absorbed by protons, that thereby transform themselves into neutrons with the emission of electron neutrinos. Thus, the star is composed mostly of neutrons, with a small amount of protons and electrons, and possibly a core made up of partially free quarks and gluons. Like electrons, neutrons are fermions, and the star is kept stable against further contraction under gravity by neutron degeneracy pressure. Once formed, neutron stars in general rotate slowly or fast, depending by the physical properties and the dynamics of their progenitor. Some of them rotate very fast, even performing a complete rotation in a few milliseconds. They generally have a strong magnetic field whose axis is misaligned with the axis of rotation, and along the north-south directions of which two powerful jets of relativistic particles and radiation are emitted. Thus, these jets precede about the axis of rotation of the star. In a few cases, the Earth happens to lie close to or on the cone of precession, hence receiving regular pulses of radiation, one pulse per rotation. In these cases, the neutron star behaves for us like a lighthouse, and is called a \emph{pulsar}. Isolated neutron stars that are not pulsars and that are not newly formed are not seen, hence they contribute to the missing mass.\\
If the compact remnants of a supernova has a mass that abundantly exceeds two solar masses, even neutron degeneracy pressure cannot balance the pull of gravity, and a \emph{stellar black hole} is formed.\footnote{\label{foot17}Being a solution of Einstein's equations of general relativity, a black hole is a classical (i.e. non quantized) object. Its gravitational field is so strong that anything that falls into the hole is lost forever inside and cannot be retrieved. This applies also to the electromagnetic field (from gamma rays to light, to radio waves): if an electromagnetic pulse of any frequency falls into a black hole, it gets permanently trapped inside and cannot leak out again. The term black hole was coined by John A. Wheeler and we understand the reason why: to anybody looking at an isolated black hole, the hole would appear completely black, as radiation of any kind cannot be emitted by the black hole. But beware! This is true only classically. In two celebrated papers \cite{explosions, 1976}, Stephen Hawking exploited the fluctuations of quantum fields near the surface of no return of a black hole (the so called black hole's horizon), to prove that a black hole emits thermal radiation. This radiation has gone down in history by the name of Hawking radiation. The temperature of Hawking radiation is inversely proportional to the black hole mass and even for stellar black holes it is extremely low: a black hole of one solar mass has a temperature of only 60 nanoKelvins (60 billionth of a Kelvin). Therefore, the only chances of observing black holes evaporate by losing energy by Hawking radiation, would be if there existed black holes of very small mass. We will shortly take up this issue in the main text.} Larger black holes, with masses ranging from, say, about one hundred to one million solar masses are thought to form through successive mergers of stellar mass black holes. In this range of masses, one speaks of \emph{intermediate mass black holes}. There exist also black holes having masses of several millions and even billions of solar masses. These monsters are termed \emph{supermassive black holes} (SMBH). Observational evidence indicate that every large galaxy has a supermassive black hole at its centre.\footnote{\label{foot18}Our own galaxy, the Milky Way, hosts at its centre a supermassive black hole, Sagittarius $A^*$, having a mass equal to 4.3 million times the mass of the Sun.}
Isolated stellar mass and intermediate mass black holes contribute for sure to the missing mass.\\
To summarise: all objects listed in the present subsection are optimal candidates to account for at least part of the missing mass since, except in some particular cases such, e.g., pulsars, they either do not emit electromagnetic radiation or, for various reasons, we do not observe the radiation that they emit. However, to establish the percentage of the missing mass they could be held responsible for, we should be able to estimate their density. In the case of the Milky Way, of the Magellanic Clouds and of our closest neighbouring galaxy, the Andromeda galaxy, a method to observe massive non luminous objects consists in exploiting the effect of gravitational lensing. For stars and planets this effect is obviously much weaker than the one generated by galaxies or galaxy clusters, hence it is called \emph{microlensing}. The transit of a MACHO in front of a background object like a star or a quasar leads to two distorted images of the lensed object or to light magnification of the latter. In general, these effects allow us to calculate the mass of the lens. So far, less than a thousand microlensing events attributable to MACHOs have been spotted in the Milky Way by the Hubble Space Telescope, by Gaia and by other observatories. The statistics is not high but, binding it to the estimated probability of such alignments to take place, it has led to the conclusion that not more than 20\% of the mass content of the dark matter halo of our galaxy is attributable to MACHOs. There are reasons to believe that in elliptical galaxies this figure might be higher, perhaps up to 50\% \cite{pavel}.\\
The last class of MACHO objects that has been suggested as a possible candidate for the missing mass is the one formed by \emph{primordial black holes}, namely hypothetical black holes that could have been created in the collapse of extremely dense local regions during the inflationary era or during the early radiation dominated universe. The existence of such objects was first hypothesised in 1966 by Yakov Zel'dovich and Igor Novikov \cite{novikov}, and later further developed in 1971 by Stephen Hawking \cite{hawking}. To contribute to, or even to explain all the dark matter content of the universe, these objects should not have had masses lower than $10^{11}Kg$ at the time of their creation, otherwise they would have all evaporated as of today, due to the Hawking effect (see footnote \ref{foot17}). At any rate, to be a good candidate for dark matter, they should have a sufficiently low Hawking temperature in order to be relatively stable and almost collisionless. To date (2024), in spite of many searches, no evidence of the existence of such objects has been found. Indeed, if primordial black holes were actually created during some very early epoch by density fluctuations, one would expect them to have had a rather wide mass spectrum so that some of them would be in the last stage of evaporation today, hence rendering them observable. But nothing has been seen. 

\subsection{Cold dark matter: WIMPs}\label{DM:2.3.4}
A possible alternative, or more plausibly a complement to MACHOs, is that the missing mass be made up of massive elementary particles, to justify the fact that they would represent cold dark matter (CDM). It is clear that, if they exist, these particles, being invisible, are not subject to the electromagnetic interaction. They would be subject only to gravitation and to the weak nuclear force, which means that the probability that they interact through other types of forces is practically non existent. This probability is measured by the so called \emph{cross section},\footnote{\label{foot19}In physics, the cross section is a quantity employed to describe the probability that a particle, or a beam of particles, hitting some target, would interact with the target being thereby diffused or absorbed. It is usually expressed in $cm^2$ and, in a particle-particle interaction, it is a measure of the area of the target that is sensitive to the interaction. For example, in nuclear interactions, cross sections are typically of the order of $10^{-24}$ centimetres square ($cm^2$), which is readily understandable, since the diameter of an atomic nucleus is typically of the order of $10^{-12} cm$ (one trillionth of a centimetre).} which therefore has to be extremely small relatively to the interaction with any type of known (or unknown) particle. A possible suggestion comes directly from the Standard Model, since we know of particles that have an extremely small cross section in any known process: neutrinos. They interact only weakly. However, neutrinos have extremely small masses, hence they cannot represent cold dark matter. Therefore, the main idea is to hypothesise the existence of particles similar to neutrinos, but having a large mass. These hypothetical particles have been called WIMPs, acronym for Weakly Interacting Massive Particles.\footnote{\label{foot20}This is just the simplest idea, but nothing prevents things to be more complex. For example, we could imagine the existence of particles endowed only of colour charges (the charges of chromodynamics), that therefore, like quarks, live only in confined states (of neutral colour), forming themselves neutral massive particles, interacting only gravitationally, or through strong interactions but with extremely small cross sections. Or any other possibility entailing an extremely low probability of nongravitational interaction. Plenty of room to indulge oneself!}
This hypothesis can explain the observations, but then one has to verify whether these particles really exist or not. There are essentially two possible methodologies to find this proof: intercept somehow these particles by having them interact with some measuring instrument, or creating them in the lab. In both cases there is an important difficulty making the enterprise highly nontrivial: one must have some information on the means of interaction of such particles with baryonic matter, in order to understand how to intercept them or create them. And it is obviously here that the theoretical model becomes important.\\
To clarify the situation, it is useful to recall how neutrinos where discovered, taking tritium decay as an example. Hydrogen has three isotopes, normal hydrogen, denoted $H$, whose nucleus is made up of just one proton, deuterium, denoted $D$, whose nucleus contains one proton and one neutron, and tritium, denoted $T$, whose nucleus is composed of one proton and two neutrons.\footnote{\label{foot21}See footnote \ref{foot14}} Whereas hydrogen and deuterium are stable isotopes, tritium is unstable. It has a half-life of $12.3$ years and decays in helium three (${}^3_2He$) plus an electron and an electron antineutrino: $T\rightarrow {}^3_2He+e^-+\bar \nu_e$. The decay of tritium can be attributed to the decay of one of the neutrons in its nucleus. Indeed, free neutrons are unstable, a neutron has a half life of 13.3 minutes and decays into a proton plus an electron and an electron antineutrino: $n\rightarrow p+e^-+\bar \nu_e$ (note that, inside atomic nuclei, neutrons have longer half-lives than when they are free, and are often even stable; for example, the neutron inside the nucleus of deuterium is stable).\\
Now, in 1930, the neutron had not been discovered yet. It was discovered in 1932 \cite{chadwick}. 
What was known at that time was that some nuclei were unstable and were decaying by emitting a beta particle (later identified to be an electron or a positron), turning into a nucleus of similar mass, but whose charge was now either one unit less or one unit more. The problem was that the sum of the energy of the final nucleus plus the energy of the beta particle was less than the energy of the parent nucleus. Therefore, it appeared for the first time that the principle of energy conservation was violated. While Niels Bohr was unwillingly ready to accept this fact, a possible solution to the problem was proposed by Wolfgang Pauli. In a famous letter of December 4, 1930, written from the ETH in Zurich to the physicists assembled for a conference in T\"ubingen, in which he addressed the participants as ``Liebe radioaktive Damen und Herren (Dear radioactive Ladies and Gentlemen)'', he suggested that the conservation of energy could be restored by adding to the decay product an extra neutral particle, very light, that he called neutron, that would carry the missing energy (after the discovery of the neutron, this particle was renamed neutrino\footnote{\label{foot22}It seems that this name was jokingly suggested to Fermi by Edoardo Amaldi in 1932, during one of the meetings of the famous via Panisperna boys.} by Enrico Fermi in 1934)\footnote{\label{foot23}At that time Pauli was 30. We transcribe the very interesting conclusion of the letter: ``\ldots Leider kann ich nicht pers\"onlich in T\"ubingen erscheinen da ich infolge eines in der Nacht vom 6 zum 7 Dez. in Z\"urich stattfinden Balles hier unabk\"ommlich bin. Mit vielen Gr\"ussen an Euch\ldots gez(eichnet) W. Pauli (\ldots Unfortunately, I cannot personally appear in T\"ubingen since I am indispensable here in Zurich because of a ball on the night from December 6 to 7. With my best regards to you \ldots signed W. Pauli'').}.\footnote{\label{foot24}Actually, when later the electroweak theory was formulated, unifying at sufficiently high energy the electromagnetic and the weak interactions, it was realised that two new \emph{quantum numbers} are conserved in a weak process" the baryon number and the lepton number, these numbers being additive, like the electric charge. To a proton and a neutron, they being baryons made up of quarks, one associates the baryon number 1 and the lepton number 0, since they do not contain any lepton. Instead, to the electron one associates lepton number 1 and baryon number 0. Therefore, the only way for things to be made right, namely in order that the total initial leptonic and baryonic numbers should not change after the decay one must assign a leptonic number $-1$ to the electronic antineutrino and hence a leptonic number 1 to the neutrino. Similarly, the positron $e^+$ (the antielectron) has leptonic number $-1$. In other words, in changing from a particle to its antiparticle the leptonic number changes sign, like the electron charge does. With this terminology, the particle emitted together with the electron, for example in the decay of tritium or of a neutron, is an antineutrino.} Pauli was right but neutrinos are interacting so weakly with matter that it took 26 years before they would be experimentally detected.\\
Indeed, Pauli's guess was not a discovery, but only a reasonable theoretical hypothesis compatible with the principle of energy conservation. It was necessary to prove that a particle with this property did actually exist, showing that it appears in all other situations foreseen by the theoretical model that describes it. This took place only in 1956, thanks to an experiment performed by the American physicists Clyde Cowan and Fred Reines \cite{cowan reines}. As stated earlier, the neutron decays as $n\rightarrow p+e^-+\bar \nu_e$. This decay allows for an \emph{inverse process}, $p+\bar\nu_e\rightarrow n+e^+$, in which an electronic antineutrino hits against a proton generating a neutron plus a positron in a process of charge exchange. Now, the idea was the following. Cowan and Reines used a nuclear reactor producing a large number of neutrons, unstable, that, if the theory was correct, would produce a large number of antineutrinos with a precise distribution of decays per second, calculated by the theory. In order to generate the aforementioned inverse process it was necessary to have the antineutrinos hit a large number of protons. Hence they placed next to the reactor a tank containing 200 litres of water mixed with Cadmium chloride. The protons in the nuclei of the water molecules are optimal targets for the hypothetical antineutrinos generated by the reactor. Since the corresponding cross section is extremely small, it was necessary to have a huge number of protons. This is the case for 200 litres of water, which contain approximately $6.7\times 10^{27}$ (6.7 times one billion billion billion) molecules. Apart from the competing process of elastic scattering $\bar \nu_e+p\rightarrow \bar \nu_e+p$, when an antineutrino hits a proton, it can theoretically generate a neutron and a positron. Very soon the positron meets an electron and they annihilate each other producing a pair of gamma ray photons, each with an energy of approximately $511KeV$ ($1KeV=$ one thousand electronvolts). They could be easily recognised proving, by measuring them, that a positron has been created. Instead, a Cadmium nucleus absorbs a neutron, whereby it moves to an excited state from which, by de-excitation, it emits a gamma ray with a precise frequency. If these photons are simultaneously measured one infers that a pair $n+e^+$ has been created. The theory allows to exactly calculate how many of these processes (hence of photons) would be generated if neutrinos were actually be generated in neutron decays. In the experiment, the photon triplets were indeed detected in the amount predicted by the theory, which confirmed the existence of neutrinos: this has been the true discovery of Pauli's neutrino, the experimental confirmation of its existence.\\
Obviously, one would like to perform something similar to prove or disprove the existence of WIMPs. Here, the main problem is that, in contrast with the neutrino case, we do not have sufficient hints to construct a theoretical model, we do not know neither the mass not the cross section of such particles hitting baryons, and which reactions these collisions might originate. The only things we know is that the mass of WIMPs cannot be small\footnote{actually, in Subsection \ref{DM:2.3.5} we will see that there can be exceptions to this rule.} and that they should interact weakly.\\
Experiments for the search of WIMPs are operating or in project in many countries in the world. They are mostly set up deep underground in old mines. Indeed, operating at such depths, one mile or more underground, the upper layers of rock and dirt block practically all cosmic rays coming from space, allowing only neutrinos and WIMPs to pass through. The idea is to set up in the mines, under cryogenic conditions, large tanks of liquid scintillators like Xenon, Argon and others, hoping for some WIMPs to hit the atoms of the liquid material. In any such hypothetical event, the scattered atom of material would produce in the medium a scintillation signal whose light would be recorded by surrounding arrays of photomultipliers. We list here the most important of such facilities: LUX-ZEPELIN Experiment (US; liquid xenon); SNOLAB (Canada; argon and Charged Coupled Devices [CCDs]); XENONnT experiment (Italy, Gran Sasso Laboratories; xenon); CDEX (China; germanium detectors and thallium doped sodium iodide scintillators); PANDA-4T (China; xenon); SUPL-SABRE (Australia; thallium doped sodium iodide scintillators).\\
So far, none of these experiments has detected any signal that could be interpreted as due to some WIMP. They have been able, though, to exclude WIMP-nucleon cross sections higher than a few units times $10^{-48}cm^2$, for reasonable WIMP energies of the order of tens of $GeV$. Such limits are close to the lowest theoretical cross section values for weak interactions. This means that, were we in the future keep on detecting nothing and finally overcome these limits, we would be forced to conclude that WIMPs, if they do exist, do not interact weakly but only gravitationally, in which case they should be better called GIMPs (Gravitationally Interacting Massive Particles). However, gravitational interactions are at least 20 orders of magnitude weaker than weak interactions, which means that the corresponding GIMP-nucleon effective cross section would not be larger than about $10^{-70}cm^2$.\footnote{Actually, gravitational forces are long range, which means that their cross section is infinite. Indeed, even if a particle were moving along a path passing very far from a body attracting it gravitationally, it would still feel the latter's influence. However, in searching for WIMPs, one is looking for processes of the type WIMP$+p\rightarrow n+e^++\nu_e$ and/or WIMP$+n\rightarrow p+e^-+\bar\nu_e$, which would be detectable through scintillations cast in the apparatus by the electron and/or the positron. Now, for such reactions to take place, a WIMP should pass at distances from an atomic nucleus not greater than $10^{-12}\sim10^{-13} cm$, otherwise no interaction would take place. Hence, the effective gravitational cross section for any such process would be to all effects infinitesimally small.} In this sense there would be absolutely no chance to ever detect any collision of a GIMP against a nucleon and, would we still believe in the existence of cold dark matter, we should accept the unsatisfactory fact that we could never be in the condition to detect it directly. In other words, in the lack of acceptable explanations, we should be content with the fact that dark matter would reveal its presence only through its indirect gravitational effect on our familiar baryonic matter, its nature remaining thereby utterly mysterious.\\
However, if one keeps believing in WIMPs in spite of their elusiveness, another possibility would be to try to create them in the lab, by means of processes akin to beta decay.\footnote{\label{foot25}There exist three types of radioactive decay: alpha, beta and gamma. Alpha decay is the process whereby an unstable atomic nucleus transforms itself into a lighter one, having a \emph{mass number} (the number of protons plus the number of neutrons) reduced by four and an atomic number (the number of protons) reduced by two, plus a so called alpha particle (denoted $\alpha$), namely the nucleus of a Helium-4 atom (${}^4_2He$), which is formed by two protons and two neutrons. A typical example is the decay of ${}^{238}_{\ 92}U$ (uranium-238) into ${}^{234}_{\ 90}Th$ (thorium-234): ${}^{238}_{\ 92}U\rightarrow {}^{234}_{\ 90}Th+\alpha$. Beta decay is a typically weak nuclear process in which an unstable atomic nucleus decays by either emitting an electron plus an electron antineutrino, or a positron plus an electron neutrino. The first case corresponds to a neutron in the parent nucleus undergoing the process $n\rightarrow p+e^-+\bar \nu_e$. The second case to a proton in the parent nucleus undergoing the process $p\rightarrow n+e^++\nu_e$ (note that a free proton could not undergo such a process since its mass is a bit less than the mass of the neutron, hence the process would not conserve energy). Electrons and positrons are called beta-particles ($\beta$-particles). Finally, gamma decay ($\gamma$-decay) is the electromagnetic process whereby an atomic nucleus in an excited state emits an energetic photon (a gamma ray) and acquires a more stable state.} Since they are expected to be quite massive, much energy would be needed to create them, and one would imagine processes of this kind to be feasible in large accelerators (like the LHC in Geneva), where very heavy particles are already produced by having very energetic beams of protons collide. However, the particles seen so far in such collisions cannot be the mysterious WIMPs, since they are unstable and decay into lighter particles into tiny fractions of a second. We also do not know what energy would be required to generate WIMPs, since we don't know their mass. Furthermore, since their cross section for interaction with ordinary matter should be very small, we would expect to create few of them anyway, even in the presence of a very large number of collisions. Finally, we do not even know what would be the right particles to make collide, in order to produce our WIMPs, except that they should interact weakly. The precise rules with which particles are produced depend on the theoretical model which describe them. But for WIMPs we do not have any model.\\
As an example of the importance of having a theoretical model, consider an electron-positron collision, in which the electromagnetic interaction plays the dominant role. Many outcomes are possible. Among them we may have an elastic scattering, a production of a muon-antimuon pair, or an annihilation into a pair of gamma photons. The cross sections for these processes measures the probability for each of them, and if the energy of the two colliding particles is very small, the cross section for annihilation in two photons is the dominant one, whereas the one for production of a muon-antimuon pair is zero, otherwise it would violate energy conservation. On the other hand, if the collision takes place at energies larger than twice the rest mass of the muon, the cross section for the $\mu^-\mu^+$ production switches on and starts growing, but it decreases rapidly when the energy exceeds three times the muon rest mass; muon-antimuon pairs are created only within a restricted energy range. All this is justified by the Standard Model  and the example shows how the production of particles depend on their properties. Hence, how can we make reasonable hypotheses regarding masses and cross sections of WIMPs? One possibility is to hypothesise some model for the production of WIMPs, in which their mass appears as a free parameter. For example, we may try to describe them with the same equations as those for neutrinos, but imposing a large mass, and then devise experiments on the basis of such model, trying to predict their outcomes as a function of the hypothetical mass. Other possibilities are suggested by theories trying to generalise the Standard Model. An example, which flourished already starting from the sixties of last century, and that for a while looked very promising, is the one of supersymmetric theories. These theories \cite{weinberg super}, that we refrain from discussing here,\footnote{\label{foot26}It suffices to say that supersymmetry predicted the existence, for each known elementary particle, of a so called supersymmetric partner.} predict the existence of several new particles,  that however have never been seen in spite of the many attempts to detect them with the great accelerators. Among them there is one, called \emph{neutralino}, that had several ideal characteristics to be a good WIMP candidate. In the simplest case (there exist various supersymmetric theories) one predicts the existence of 4 different types of neutralinos, all massive and interacting only weakly, of which one alone is stable.\footnote{\label{foot27}If WIMPs exist, they must necessarily be stable particles, since cold dark matter appears to have always existed.} The advantage of this approach is that the supersymmetric theory provides precise predictions for the mass, possible interactions and corresponding cross sections for the neutralino. And the calculation made in the nineties seemed to indicate that neutralinos were perfect candidates to explain dark matter. Their cross section was sufficiently high to allow for many of them to be produced during the Big Bang. In addition, its mass multiplied by the expected number of created neutralinos, did provide the correct amount of dark matter assumed to exist. Two different areas of physics, on one side particle theory, on the other side astrophysics and cosmology, were providing a coherent description of the quantity of matter in the universe, with different independent arguments. This exciting coincidence was termed the ``supersymmetric miracle''.\\
Unfortunately, however, all numerous experiments, devised and performed to verify this hypothesis, have instead led to the conclusion that the neutralino does not exist. As stated earlier, many other experiments are in progress or under planning with the aim to try to detect WIMPs of a different hypothetical nature (see also the last section of the Chapter on dark energy). To date, all these experiments have turned out to be inconclusive. So, the fact that no WIMP has been detected yet, leads to the conclusion that the cross sections for the production of these particles in the various foreseeable processes is much smaller than any one hypothesised so far. This leads to put a supplementary question: if the cross section for WIMP production is so small, it is difficult to believe that, at the time of the Big Bang, so many WIMPs could have been produced compared to the particles of the Standard Model. Thus it follows that they should be overly massive so that the hope of being able to produce them in an accelerator vanishes. Therefore, even though the WIMP hypothesis cannot be entirely excluded, it keeps becoming less plausible as time goes by, It is therefore natural to push for research in alternative directions to give more convincing justifications of the effect of the missing mass.

\subsection{Dark matter as axions or axion-like particles?}\label{DM:2.3.5}
There are other candidates that have been proposed to explain dark matter. One of them is the \emph{axion}, a particle still hypothetical to date, whose existence, however, would lead to a simple solution of a shortcoming of the Standard Model: the \emph{strong CP problem}. Symmetries have always been a crucial guide in the comprehension of physics. However, some symmetries may be relevant only in certain situations and not others. In the Standard Model, besides relativistic invariance, which is required by special relativity, there are three important discrete symmetries: time reversal (T)\footnote{\label{foot28}Recalling the metaphor in Subsection 1.3 of the Chapter on dark energy about throwing a rock vertically upwards from the surface of the Moon, suppose we are again standing on the surface of our satellite and we throw up a marble ball in some given direction, with a certain initial speed. Following Newtons's law of gravity, the ball will travel along an arc of a parabola and eventually fall on the ground. However, suppose instead that at a certain instant, when the ball is still in mid air, we were able to invert the direction of its velocity, without changing the magnitude of the latter. As a consequence of such an action, the ball would simply retrace back the same path that it had followed before its velocity had been reversed. In other words, it would move as if we were projecting backward the movie of its motion: everything would unfold as if the direction of time had been reversed, namely if the ball were traveling backward in time.\\
Another, more complex but more illuminating example is the following. Consider an ideal billiard table with rounded corners: no holes, no friction, elastic bumps between balls and upon hitting the walls. Start with an initial situation in which all the balls are at rest with the exclusion of one of them being thrown, from its initial position, in a given direction with a given velocity. After a while and several bumps, the balls will appear distributed on the table more or less randomly and having random velocities. However, if, at a given instant, we would be able to reverse the velocities of the balls, they would all retrace back their motion and bumps, and eventually return to their initial configuration. In this example as well the whole motion unfolds as if time has been flowing backwards. These are examples of the operation of time reversal.}, charge conjugation (C) and parity (P). T changes the sign of the flow of time, and it is a symmetry if the laws of physics do not depend on the direction of time flow. C transforms particles in their corresponding antiparticles and it is a symmetry if, from the point of view of its dynamical laws, a world of antiparticles is entirely equivalent to a world of particles. P transforms any dynamical process in its reflected in a mirror. And, again, we assert P to be a symmetry if the laws of physics do not change under such reflection. These transformations are for sure all symmetries of electrodynamics: no process mediated by the electromagnetic interaction (such as for example electron-positron annihilation $e^-+e^+\rightarrow \gamma+\gamma$) violates any of the three. However, in 1956 it was found that P, parity, is violated in weak interactions. The Chinese-American physicists Tsung-Dao Lee and Chen-Ning Yang, after a careful analysis realised that, whereas P was certainly a symmetry in all experiments involving electromagnetic and strong interactions, nevertheless neither the theory nor any particle interaction experiment performed up to that time, would necessarily imply P to be a symmetry. Therefore, they proposed some experiments devised to test whether parity was conserved or violated in weak interactions \cite{lee yang}.\footnote{\label{foot29}The fact that the existence of a symmetry implies, and is implied, by the conservation of some physical quantity is a well known fact in physics. In particular, in the case of continuous symmetries, it is the content of the celebrated Noether's theorem \cite{noether}} One of these experiments was performed in the lab by the Chinese-American physicist Chien-Shiung Wu (or Jianxiong Wu, also known as Madame Wu), and confirmed that parity was violated \cite{wu}. In the same year it was also proved that parity violation in weak interactions was also violating C symmetry \cite{ioffe}. On the other hand, in 1957, the Soviet physicist Lev Davidovich Landau, proposed that, in spite of these violations, the combined symmetry CP would still be universally preserved \cite{landau}. However that eventually turned out to be a wishful thinking. Indeed, in 1964 the American physicists James Cronin and Val Fitch proved that CP was violated in the weak decay of certain neutral mesons, termed K-mesons or kaons \cite{cronin}. In the following years, the Cronin-Fitch result was confirmed in many other types of weak meson decays. Thus, not only P, but CP as well, is violated in weak interactions, and these violations are well described by the Standard Model in electroweak theories. \\
But what about strong interactions? As we have seen in Subsection 1.3.1, the strong interactions are well described in the Standard Model by the theory termed quantum chromodynamics (QCD), in which the strong force between quarks is transmitted by zero mass self-interacting particles called gluons. According to QCD, a violation of CP symmetry in strong interactions could occur, but it has never been observed in any experiment involving the strong interaction alone. This hitherto unexplained and puzzling fact been termed the strong CP problem. So what prevents this violation? In 1977, the Italian-Argentinian physicist Roberto Peccei, and the Australian physicist Helen Quinn observed that if one hypothesises the existence of a quantum (pseudo)scalar field coupled in some specific way to the chromodynamical fields, then a new symmetry (successively called Peccei-Quinn symmetry) would prevent the violation of CP in the strong interactions \cite{peccei, quinn}. However, this symmetry is spontaneously broken by the dynamics,\footnote{\label{foot30}Generally speaking, spontaneous symmetry breaking occurs in a physical system, whenever the dynamical equations of motion of the system display some symmetry (for example the symmetry under rotations in space), but the lowest energy state into which the system has the tendency to spontaneously fall, lacks the symmetry.} thus generating a possible small CP violation. The physicists Frank Wilczek \cite{wilczek} and Steven Weinberg \cite{axion} proved that this spontaneous symmetry breaking implies the existence of a neutral particle having spin zero and a very small mass, less than $10^{-2} eV$ (one hundredth of electronvolt), that couples weakly with matter, and that they called \emph{axion}.\footnote{\label{foot31}There exists an important theorem, proved in the context of special relativity, called the CPT theorem, that states that CPT is a universal symmetry. Hence CP violation implies that, at least in weak interactions, T is violated as well.} Unlike neutrinos, such a small mass would not a priori prevent these particles to be good candidates for cold dark matter. The reason is that axions are bosons. Hence, unlike neutrinos that are fermions, they are not subject to the Pauli exclusion principle (see footnote \ref{foot15}). Therefore, at very low temperatures they can assume a state, called \emph{Bose condensate}, in which they all fall into the lowest energy state and can thus accumulate in spatially limited regions. The idea is that the expansion due to inflation would have suddenly cooled down the primordial axions, bringing them in a condensed state, forming in this way dark matter halos by a quantum effect instead than by a gravitational one. There is yet no proof of the existence of axions, and the most recent measurements that their mass cannot exceed $10^{-6}eV$ (one millionth of electronvolt), too small to have cosmological consequences, does not however rule out the possibility that these particles could still be relevant for the dynamics of galaxies. Other very light hypothetical particles, akin to axions, have been proposed as a possible explanations of cold dark matter. They have been dubbed Axion-Like-Particles (ALP). Their crucial characteristics is that they are assumed to be self-interacting. This property is described by nonlinear equations that allow for the existence of localised stable configurations, similar to electromagnetic solitons in nonlinear media, or even to solitons observed on the surface of shallow waters \cite{shallow}. Galactic dark matter halos could be ALP solitons. Experiments devised to find ALPs consists in trying generating them through the \emph{Primakoff effect} \cite{primakov}, according to which photon beams that cross intense electromagnetic fields could, among other things, be converted into axions and/or ALPs and vice versa. Based on these processes are experiments like ADMX (Axion Dark Matter eXperiment) at the University of Washington in Seattle and XENONnT at the Gran Sasso Laboratories, Italy.\\
Encouraging results for ALPs come from the study of the gravitational lensing effect of the galactic system HS 0810-2554 \cite{amrut 2023}. This object is an elliptical galaxy at $z=0.89$ that gravitationally lenses a background quasar at $z=1.51$, generating four distorted images of the latter. The structure of the dark matter halo of the foreground galaxy, inferred from the distorsion of the images of the background object, shows more dense and less dense regions that can be reasonably interpreted as the result of an interference pattern generated by a quantum mechanical wave soliton of a large aggregate of ALP particles with masses expected between $10^{-22}$ and $10^{-20}eV$. This example, which is supported by theoretical models, seems to favour dark matter halos formed by ALP against those formed by WIMPs.\\
Of course, before reaching definite conclusions in favour of dark matter ALPs, one should be able to observe other examples of lensing galaxies displaying analogous features. It has been noted \cite{bird 2023} that ALP fields may also play a role in explaining the very early formation of massive galaxies observed by the JWST \cite{jwst, z11}. Indeed, the interaction of ALPs with other particles is assumed to be too weak for ALPs to reach thermal equilibrium with the rest of the early universe plasma. This would then lead to oscillations of the ALP field which, within a large parameter range of ALP masses ($10^{-22}eV \lesssim m_{alp}\lesssim10^{-19}eV$), would allow for subsequent fragmentations of the ALP field into massive ALP clusters acting as potential wells for the formation of such massive early galaxies.\footnote{One should observe that in \cite{mack} the American astrophysicist Katherine Mack has shown that, even in the case that the mystery of dark matter were solved, the existence of ALPs would introduce several secondary problems that it would be very difficult to solve, even with the introduction of an anthropic principle.}

\subsection{No dark matter: modified gravity theories}\label{DM:2.3.6}
The failure so far to detect the constituents of dark matter has convinced a large minority of physicists and astronomers to put forward the hypothesis that dark matter doesn't exist at all, its effect being illusory, due to the fact that the gravitational dynamics at large scales is no more described by Newton's laws. Indeed, all phenomena that have led us to postulate the existence of dark matter are purely gravitational in nature. The other type of forces would come into play only as a consequence of the true existence of such matter. Therefore, we are justified in asking ourselves if the fact that we cannot make ends meet, doesn't mean that there is extra matter that we are unable to see, but rather that we are using the wrong formulas to describe the dynamics on scales much bigger than those at which we had verified it so far. It is then obvious that we should understand which modifications we should make to the laws of gravitation in order for these to correctly explain the dynamics of galaxies and of larger structures, without the need to add extra unseen matter, but still preserving the old dynamics at small scales. Theories obtained in this way are called MOG (Modified Gravity). We will display the leading ones, with their pros and cons.

\subsubsection{MOND}

\begin{figure}
\label{Figure4}
	\begin{minipage}[b]{1.0\linewidth}
		\centering
		\includegraphics[width=8 cm]{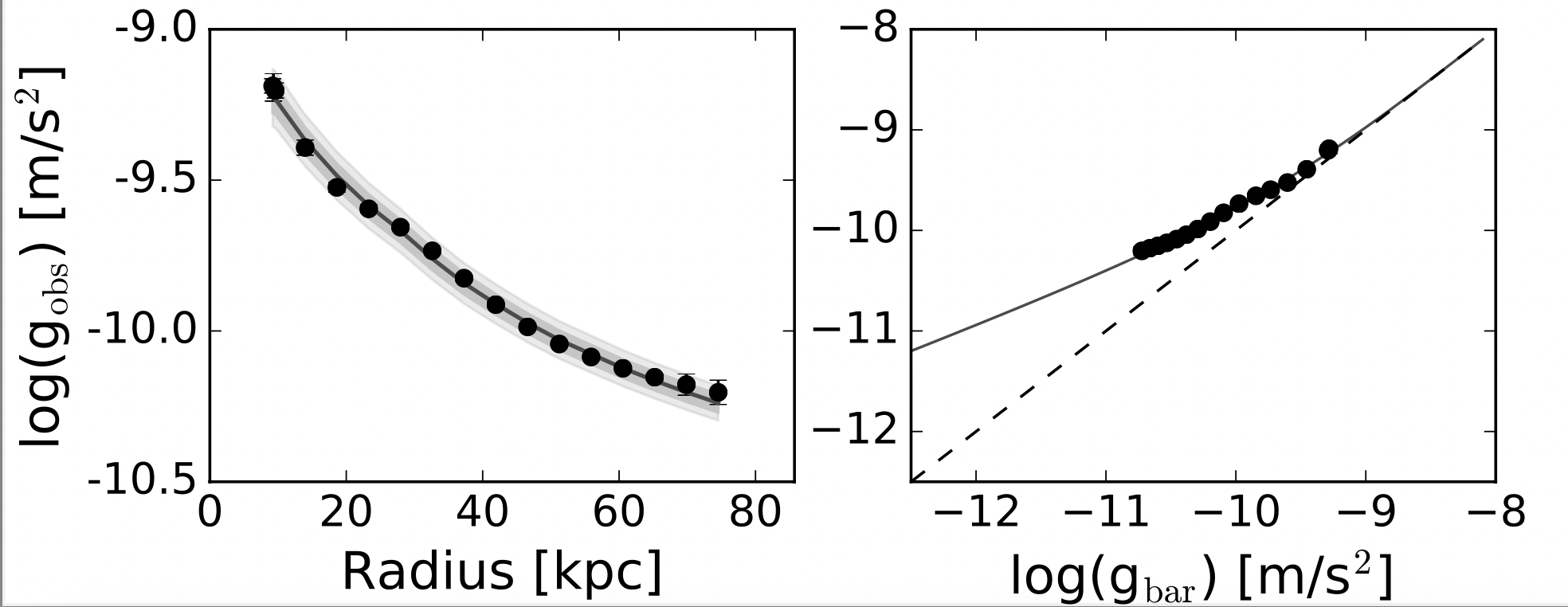}
		\caption{\label{1.4} Graph of the Newtonian gravity acceleration vs the observed acceleration for the galaxy UGC02487 (from the SPARC catalogue, \cite{lella}). Axes are in logarithmic scale.}
	\end{minipage}
\end{figure}

In order to understand how dynamics changes, when assuming that there is no dark matter, we may proceed as follows. Giving any mass distribution that acts on a body orbiting around it, we use Newton's law of gravitation to calculate the theoretical acceleration $g_{bar}$ that the visible mass (the baryonic mass) exerts on the body. At the same time one measures its observed acceleration $g_{obs}$. By repeating such measurements on different systems, we can display the data pairs $(g_{obs},g_{bar})$ on a graph to verify the relation between the two, called RAR (Radial Acceleration Relation). See Fig. \ref{1.4}.
If Newton's law were always correct, the graph should be a straight line bisecting the represented quadrant. From the figure we see that $g_{obs}\simeq g_{bar}$ for accelerations larger than approximately $10^{-9}$ metres per square second, only to diverge from the straight line for accelerations below that value. The discrepancy from the straight line seems to be compatible with the relation $g_{obs}\simeq \sqrt {a_0 g_{bar}}$, for $g_{bar}$ much smaller than $a_0$, where $a_0$ is of the order of $10^{-10}$ metres per square second. Therefore, if we want to modify the law of gravity, with the hypothesis of absence of dark matter, the modifications introduced should be coherent with these observed data.
In 1983, the Israeli physicist Mordehai Milgrom \cite{milgrom 1, milgrom 2, milgrom 3} proposed a change of Newton's law by asking that the acceleration $g$ of a test particle subject to the gravitational field generated by a point mass $M_b$ placed at a distance $r$ be given by the formula (compare footnote 15 of the Chapter on dark energy) $GM_b/r^2=\mu(g/a_0)g$, where $\mu(x)$ is a suitable interpolating function, to be chosen as follows. For accelerations much larger than $a_0$, the factor $\mu$ should be approximatively equal to 1, in accordance with the conventional Newton's law. For smaller accelerations, the behaviour of $\mu$ should be determined so as to reproduce the graph in Fig. \ref{1.4}. We obtain the empirical law $g_{obs}=\sqrt{a_0 g_{bar}}$ by taking $\mu(g/a_0)=g/a_0$. Therefore, the Milgrom correction requires choosing for $\mu(x)$ a function approaching the value 1 for $x$ large, and approaching $x$ for $x$ small. Fore example, the graph of Fig. \ref{1.4} is well interpolated by the choice $\mu(x)=x/(1+x)$. The commonly accepted value for $a_0$ is $1.2\times 10^{-10}m/s^2$. The law so obtained is called MOND (MOdified Newtonian Dynamics).\\
The most important achievement of MOND is that it explains the rotation curves of galaxies, with their flat profile. For a circular motion of radius $r$ with constant velocity v, the centripetal acceleration is given by $g={\rm v}^2/r$. At very large distance from the centre of the galaxy the acceleration becomes very small and MOND's law takes the form $g=\sqrt{a_0g_{bar}}=\frac {\sqrt{a_0G M_b}}r$. Comparing with the formula for circular motion we obtain for the velocity the value v$=\sqrt[4]{a_0GM_b}\equiv {\rm v}_f$. It is a constant value, precisely as in the flat rotation profile that results from observations (see Subsection \ref{DM:2.2.1.1} and Fig. \ref{1.1}). Furthermore, MOND also accounts for the Tully-Fisher relation, that the $\Lambda$CDM model is unable to explain. Indeed, from the formula for the peripheral velocity, we see that the baryonic mass of the galaxy is proportional to the fourth power of v$_f$.\footnote{See section \ref{DM:2.2.1.1}.} However, note that Milgrom's law $GM_b/r^2=\mu(g/a_0)g$ is no more than a trivial consequence of the phenomenological law $g=\sqrt{a_0 g_{bar}}$, resulting from the graph of Fig. \ref{1.4}, which in turn represents simply the results of the observations of the rotation curves of large samples of disc galaxies. Therefore, if one believes that Milgrom's modification of Newton's law (and the characteristic acceleration $a_0$) has a much deeper meaning, one must show that it explains many more, and possibly all, known evidences of the existence of a missing mass phenomenon, without the need of invoking dark matter. Some prominent examples in which MOND works well are the following. \\
In disc galaxies Newton's law applies with virtually no modifications in explaining the average acceleration of stars at any radius from the centre, up to well beyond the outskirts of the galaxy, and it has to be modified only when the centripetal acceleration decreases to values close to $a_0$. This means that the purported dark matter halo is mostly concentrated in the outskirts of the galaxy, a feature for which there is no satisfactory explanation in the context of the $\Lambda$CDM model. On the other hand, this so called discrepancy-acceleration relation is neatly explained by Milgrom's law.\\
The acceleration constant $a_0$ features an upper cutoff to the mean surface brightness of galaxies, as observed and formulated in the Freeman law \cite{freeman 1970} for discs, and in the Fish law \cite{fish 1964} for elliptical. Whereas, in conventional dark matter-based galaxy formation models, this property has to be put in by hand.\\
Several ultra diffuse galaxies (UDG) \cite{udg} have dimensions comparable to those of typical disc galaxies, but they contain only about $1\%$ of the stars present in these galaxies. Many of the UDGs seem to contain more than $99\%$ dark matter, compared to the $80\%$ or so of missing mass in the universe. MOND would readily explain this. Indeed, the stars of such galaxies are so dispersed, that the average distance between neighbouring pairs is so large that their reciprocal gravitational acceleration is generally smaller than $a_0$. According to MOND, this would readily account for such a high percentage of ``missing mass'', distributed all over the galaxy.\\ 
The tidal deformations of dwarf galaxies in the Fornax cluster, and their lack of low surface brightness towards the centre of the cluster, are features difficult to explain within the $\Lambda$CDM model, whereas they are well accounted for by MOND \cite{asencio 2022}.\\
The most striking property predicted by MOND, which follows from the non-linearity in acceleration, is the so called external field effect (EFE), according to which, besides its internal dynamics, the behaviour of any system is affected by the whole universe, even at the strict local level, thus disrupting the strong equivalence principle.\footnote{The strong equivalence principle states that, relative to a free falling observer, all laws of physics are locally the same as in special relativity, namely in the absence of any gravitational field. This principle lies at the foundations of general relativity, which is why relativistic generalizations of MOND (see next subsection) differ from general relativity.} And, indeed, for example, an analysis from the Spitzer Photometry and Accurate Rotation Curves (SPARC) sample, together with estimates of the large scale external gravitational field from an all-sky galaxy catalog, led some researchers to conclude that there was highly statistical significant evidence of violations of the strong equivalence principle in weak gravitational fields in the vicinity of rotationally supported galaxies \cite{chae 2020}.\\
These results, which point to a breakdown of the strong equivalence principle, require a modification of general relativity, in order to account for the effects of the critical acceleration $a_0$. We deal with this attempts in the next subsection. There are, however, several astrophysical situations that MOND faces difficulties in explaining. We list some of them.\\
Galaxy clusters exhibit a missing mass that MOND is able to explain only partially. This has lead some researchers to suggest that in clusters there is some hitherto undiscovered non baryonic extra mass, such as perhaps some neutrino contribution \cite{angus 2007, sanders 2007, nieuwenhuizen 2016}.\\
Similarly, MOND encounters difficulties in explaining the temperature profiles of galaxy clusters and the velocity dispersion profiles of globular clusters \cite{aguirre 2001}. A difficulty for MOND is that, while some UDGs exhibit a very high percentage of missing mass, a feature that, as we stated, MOND explains well, others seem to contain very little dark matter, or none at all. In conventional dark matter explanations, this feature is justified by contemplating two possibilities. Either these UDGs originated via tidal tail instability in the vicinity of a massive host galaxy, or via high velocity collisions of dwarfs. In the second case, when two dwarfs collide, the stars and dark matter of each member of the pair pass through each other essentially unscathed, while the hot intergalactic gas and dust practically stop by attrition in the region of the collision. Then these gas and dust, left by themselves, attract some more matter by the surroundings, compactifies itself and, by local inhomogeneities, starts to form stars, eventually giving rise to an UDG with very little or no dark matter \cite{van dokkum 2022, silk 2029, mancera 2022, moreno 2022, shin}. On the other hand, there have been attempts to save MOND even in such systems, for example by invoking the external field effect (EFE) of close-by galaxies or a wrong evaluation in the observed inclination of the galaxies' discs \cite{muller 2019, banik 2022}.\\
Another difficulty encountered by MOND lies in the problem of explaining the properties of the Bullet Clusters, that, as we know (see Subsection \ref{DM:2.2.1.5}), provide probably the most impelling indication of the existence of dark matter. Indeed, in such systems, the concentration of dark matter is located off centre, on one and, respectively, on the other side of the central region, the latter being occupied by the hot gas: according to the fact that, during the collision of the two galaxies, the stars and the dark matter of each member of the colliding pair has passed through without being appreciably slowed down. Such a configuration would not be compatible with MOND, since greatest part of the baryonic mass is in the clusters' gas, and therefore the halo should be concentrated around the latter. However, some attempts have been made to save MOND even in this circumstance \cite{angus 2006}. Instead, it is conceivable that, in a few million years, the two known Bullet Clusters will have left, in the region of the collision, a dark matter-free galaxy, the final outcome of the progenitor central hot gas.\\
A few years ago a study of a sample of 193 disc galaxies was carried out \cite{rodrigues 2018} with the purpose of comparing the dark matter hypothesis of the $\Lambda$CDM model with the one by MOND replacing dark matter with the effects of a purported fundamental acceleration scale $a_0$ modifying Newton's law of gravity. In the words of the authors: ``We show that the probability of the existence of a fundamental acceleration that is common to all the galaxies [of our sample] is essentially zero. \ldots In particular, the MOND theory, or any other theory that behaves like it at galactic scales, is ruled out as a fundamental theory for galaxies at more than $10\sigma$.'' However, the claim in \cite{rodrigues 2018} was challenged in \cite{kroupa 2018, mcgaugh 2018} on the basis that the conclusions of the aforementioned authors rely on galaxies with very uncertain distances and/or nearly edge-on orientations, neglecting errors on galaxy inclination. Therefore, these authors reaffirm the validity of MOND.\\
Among the rich galaxy phenomenology are the so called massive early-type galaxies (ETGs), having masses of the order of the mass of the Milky Way. Most of these galaxies have been shaped through a two-phase process: the rapid growth of a compact core, followed by the accretion of an extended envelope through mergers. Some of them, though exceedingly rare, have avoided the second phase. They are called relic galaxies. The best relic galaxy candidate discovered to date is the lenticular galaxy NGC 1277, in the Perseus cluster \cite{comeron 2023}. The motions of the stars near its centre suggest that NGC 1277 hosts an enormous black hole with an estimated mass equal to 17 billion solar masses (see footnote \ref{foot18} for a comparison with the mass of the black hole at the centre of the Milky Way). It appears that this galaxy is almost devoid of dark matter, a fact that seems inexplicable by MOND. In the words of Ignacio Trujillo, one of the authors of reference \cite{comeron 2023}: ``Although the dark matter in a specific galaxy can be lost, a modified law of gravity must be universal, it cannot have exceptions, so that a galaxy without dark matter is a refutation of this type of alternatives to dark matter''.\\
On the other hand, McGaugh asserts \cite{mah} that NGC 1277 is such a compact lenticular, and without any gas around it, that the accelerations of all its components are never less than $2a_0$. Such being the case, the Newtonian regime would dominate, without any contradiction with MOND.\\
Can the Mondian external field effect explain the lack of missing mass in galaxy NGC 1277?\\
A very interesting approach devised to test MOND has been proposed and performed by the authors of \cite{banik 2024}. In their paper, using the third data release (DR3) of the ESA space telescope Gaia, the authors study the relative motion of 8611 pairs of stars, called wide binaries (WBs), each star of a given pair being separated by the pair companion by an average distance varying from 2 to 30 KAU,\footnote{One KAU = one thousand astronomical units. An astronomical unit is the average distance of the Earth from the Sun, and is equal to 150 million kilometers. For comparison with the WBs of the article, the average distance from the Sun of the farthest planet, Neptune, equals 30AU.} depending on the chosen pair.
At such reciprocal distances, the average relative acceleration in each pair is less than the critical MONDian value $a_0\simeq 1.2\times 10^{-10}m/s^2$. The authors introduce a parameter $\alpha_{grav}$ which, for each WB of the sample, interpolates between the Newtonian and the Milgromian prediction; the parameter taking the value 0 for pure Newtonian behaviour, and 1 for pure Milgromian. By taking into account all possible disturbing effects, including in particular the EFE, the authors reach the conclusion $\alpha_{grav}=-0.021_{-0.045}^{+0.065}$, which is fully consistent with Newtonian gravity and excludes MOND at $16\sigma$ confidence.\\
However, other authors had earlier obtained different results in the same context, supporting MOND: \cite{chae 2023, hernandez 2023}. The two schools are still hotly debating the matter. It is very difficult to resolve the issue, as the accelerations at stake are indeed extremely tiny, so that many, even slight, external disturbances can entirely mask small but crucial differences.

\subsubsection{Relativistic extensions of MOND and other alternative theories}\label{DM:2.3.3.2}
If MOND were to represent a truly fundamental theory, it should be liable to be generalised, from its original formulation as a modification of Newton's theory in the nonrelativistic domain, to a full relativistic theory, thus generalising Einstein's equations of general relativity. This is necessary, since many galactic and cosmological phenomena are relativistic in nature. The relativistic extensions of MOND are collectively called RMOND. They should be able to reproduce not just the rotational curves of galaxies, but all those phenomena that are usually ascribed to dark matter.\\
The first consistent attempt at generalising MOND to the relativistic domain was made in 2004 by Jacob Bekenstein \cite{bekenstein 2004}. It is based on a local Lagrangian\footnote{In mechanics and in field theory a Lagrangian is a particular function of the configurational coordinates of a mechanical system, and of their time derivatives or, respectively, of the field variables and of their space and time derivatives, in terms of which the equations of motion of the dynamical system in question can be obtained according to a universal rule. The name Lagrangian comes from the Italian mathematician, astronomer and mathematical physicist Giuseppe Luigi Lagrangia, successively naturalised in France as Joseph-Louis Lagrange (Turin, 1736 -- Paris, 1815). His most important work has been the treatise M\'ecanique Analytique (Analytical Mechanics), first published in 1788, offering the most comprehensive treatment of classical mechanics since Newton.} in which gravitation is mediated by a metric tensor, a unit vector field, and a scalar field, all three dynamical. These fields have given the theory the name TeVeS (Tensor--Vector--Scalar). The theory correctly has a Newtonian limit for nonrelativistic dynamics with significant acceleration, but a MOND limit when accelerations are small. Therefore, it explains the rotation curves of disc galaxies without the need of dark matter. In addition, it reproduces phenomena like gravitational lensing, the accelerated expansion of the universe, and the observations on structure formation. See: \cite{clifton 2012, skordis 2006}.\\
However, TeVeS faces several problems. We list a few. It has been shown that the height of the third peak in the power spectrum of the temperature fluctuations of the CMB (see Fig. \ref{DM:2.3}) favours cold dark matter against TeVeS with high probability \cite{slosar 2005}. It has also been shown that it is not possible to adjust the TeVeS function which controls the strength of the modification to gravity in such a way as to simultaneously fit galaxy lensing data and rotation curves \cite{mavromatos 2009}. In addition, TeVeS faces problems with the lifetime of compact objects \cite{seifert 2007}.\\
A more flexible version of TeVeS, that solves some of its problems by adding an additional scalar field, goes under the name of Bi-Scalar-Tensor-Vector gravity (BSTV) and was developed in \cite{sanders 2005}. An alternative to TeVeS and to BSTV, that also dispenses with dark matter, and that, in particular, allows for the gravitational constant $G$ to vary in space and time, is the so called Scalar--Tensor--Vector gravity (STVG) \cite{moffat 2006} STVG accounts for: the accelerating expansion of the universe; the acoustic peaks in the CMB; the matter power spectrum of the universe that is observed in the form of galaxy--galaxy correlations. It is also in good agreement with the mass profiles of galaxy clusters.\\
Finally, among the relativistic extensions of MOND, we mention a recent proposal by C. Skordis and T. Z\l o\'snik \cite{skordis 2021} which, unlike TeVeS, successfully reproduces the CMB spectrum and the matter power spectra.\\
If on the one hand the correction of theories by adding new fields helps to reproduce the desired phenomena, on the other hand this is not surprising, and not particularly satisfactory, since the introduction of new parameters obviously adds possibilities to adapt them to the requirements. The problem is that, in general, in all cases the adaptations add other incoherences with observations that require additional extensions that exceedingly complicate the models.\\
Then, many other models have been constructed, trying to minimise the introduction of new parameters (see Section 3.5 of the Chapter on dark energy). Among these are the $F(R)$ theories, first introduced by the German--Australian physicist Hans Adolf Buchdahl in 1970 \cite{buchdahl}, and Brans--Dicke's theory \cite{brans}.\\
In addition to these, there exists an innovative proposal by the Dutch physicist Erik P. Verlinde \cite{verlinde 2011} who, inspired by earlier works on the connection between gravity and thermodynamics, formulated a theory according to which gravity would not be a fundamental force, but instead an emerging phenomenon from the statistical dynamics of fields and particles. Precisely, gravity is explained as an entropic force caused by a change in the amount of information associated with the positions of bodies of matter. In this regard, the statistical average should give the usual Newton's laws. But then one also has to study the fluctuations in the gravitational force. Their size depends on the effective temperature, which is quantum mechanical in origin. Now, it is conceivable that fluctuations may turn out to be more pronounced for weak gravitational fields, thus explaining the fact that, at large distances from a body, hence for very small accelerations of test particles, the gravitational force should fall off more slowly than implied by Newton's law, thus solving the missing mass problem without resorting to ``dark matter'' as an explanation. In a relativistic context, Verlinde's approach leads to Einstein equations, as expected. Recently, the approach to entropic gravity has been further developed in \cite{schlatter 2023}. In particular, the theory also naturally gives rise to a cosmological constant and to a modified Newtonian dynamics, thus providing a physical explanation for the phenomena historically attributed to dark energy, and dark matter as well.\\
\underline{GW170817}\\
Since the time when it has been formulated \cite{Einstein 1915}, general relativity has witnessed many confirmations of its validity. Among the most recent ones is the discovery of gravitational waves, which took place the 14{th} of September 2015 \cite{gw150914}. In general relativity, matter-energy is a source of distortions of spacetime geometry (see Section 1.3 of the Chapter on dark energy). If generated by some time -dependent phenomenon, such as changing mass-energy currents, these distortions can propagate in vacuum at the speed of light, in the form of wavy oscillations. It is an analogue of electromagnetic waves: electric charges and electric currents varying in time generate an oscillating electromagnetic field, that propagates in empty space and that we call an electromagnetic wave. Similarly, a perturbation of the spacetime geometry generated by moving masses propagates through space in the form of a gravitational wave. According to Einstein equations, the generated perturbations are proportional to the mass-energy density and mass-energy current of the source multiplied by the coefficient $\kappa=8\pi G/c^4$, where $G$ is the gravitational constant and $c$ is the speed of light. On account of the smallness of $G$\footnote{$G=6.6743\times 10^{-11}m^3Kg^{-1}s^{-2}$} and of the appearance of the fourth power of $c$ at the denominator, one can easily realise that the propagating spacetime distortions will in general have a very small amplitude, unless generated by some catastrophic phenomenon such as, typically, a supernova explosion or a merging of two compact objects (black holes and/or neutron stars). Indeed, the reader can enjoy verifying by himself that $\kappa\simeq 2\times 10^{-43}m/joule$. Furthermore, while propagating, the energy of the wave spreads out on a sphere of increasing radius at the speed of light. Therefore, the transported energy arrives at a distant observer in proportion to the inverse square of the distance. The greatest part of the events that have been observed to date have been generated by the merging of pairs of black holes with masses of a few tens of solar masses, with conversion in the form of gravitational waves of the energy of a few solar masses. The gravitational signals arrived to us from distances of a few hundred billion light years have been so weak as to cause on Earth geometrical deformations of the order of fractions of the radius of a proton. And since in black holes the event horizons forbid anything to escape, practically no other signals accompany the gravitational waves in such cases.\\
On the other hand, the 17{th} of August 2017 the Earth was hit by a peculiar gravitational wave, generated by the merging of two neutron stars, that took place in the galaxy NGC 4993 located at a distance from us of about 130 million light years. Neutron stars have no horizons and, in this merging, an enormous amount of neutron rich debris was ejected, emitting, 1.7 seconds after the gravitational wave signal, a burst of gamma rays, followed, in the course of a few weeks, by electromagnetic waves distributed along the whole spectrum, from X--rays down to radio waves. These electromagnetic signals were all detected by telescopes on Earth. In the course of this event an enormous amount of neutrinos was also emitted but, being it so far away, no neutrinos originated from it have been detected on Earth. The event has been called GW170817, meaning Gravitational Waves 2017 August 17  \cite{gw170817}. It is considered the act of birth of the so called \emph{multimessenger astronomy}. Meaning the era of astronomy in which one can study astronomical objects from many different points of view, by observing their electromagnetic spectrum, the neutrino spectrum, and the gravitational one.\footnote{To date, more than 90 gravitational wave events have been detected by the astrophysicists of the collaboration working with the laser interferometer gravitational wave observatories LIGO in the US, Virgo in Italy, and KAGRA in Japan. Apart from the event GW170817 that we have described, and the three events GW191219, GW200115 and GW200210, that most probably have featured mergers of a black hole with a neutron star, the other events detected all correspond to mergers of black hole pairs, each member of the pair having mass ranging from a few tens to several tens times the mass of the Sun. The reason for this large discrepancy is that the mass of a neutron star is never much greater than two solar masses. Hence, neutron star--neutron star and black hole--neutron star mergers generate much less energy in gravitational waves than the mergers of pairs of massive black holes, and are therefore much more difficult to detect. There is much hope that further improvements in the performance of the present interferometric detectors and the construction of more advanced future ones, will allow the observation of many more events in which at least one of the merging components is a neutron star, thus considerably boosting information from multimessenger events.} This would provide very refined possibilities to verify gravitational theories. For example, GW170817 has allowed us to establish that gravitational waves propagate in empty space with the speed of light, with a 15 digits precision.\footnote{Note that this does not mean that gravitational and electromagnetic waves arrive at the Earth simultaneously. Indeed, we have seen for example that the gamma ray burst from GW170817 was detected 1.7 seconds after the observation of the gravitational wave signal. The reason for the delay is that, starting from the same merging pair, these waves are not just crossing empty space to arrive to us, but in general are also going through regions of matter. In doing so gravity, which interacts so weakly with matter, passes across with a substantially unaltered speed. Whereas light is slowed down in different manners, depending on its wavelength and the matter components crossed. The precise knowledge of the interaction between the electromagnetic field and and matter allows for a detailed analysis of the process.} This result has allowed to refute several proposed modified gravity theories that, though relativistic, foresee that gravity would propagate at speeds different from light speed \cite{skordis 2019}. \\
\underline{Binary pulsars}.\\
In general, a neutron star spins fast around its axis, from a few turns per second to sometimes more than 1000 turns per second. It also generates an intense magnetic field whose axis is in general tilted relative to the spin axis. Hence, as the star spins the axis of its magnetic field sweeps a cone whose axis is the spin axis. The magnetic field of a neutron star funnels two intense jets of particles and radiation along the magnetic poles. An isolated neutron star is invisible from Earth unless our planet happens to lie on, or close to, a generatrix of the cone spanned by the direction of the star's magnetic field. In the latter case, the Earth receives a pulse of radiation from the star's jet, which repeats itself once every turn. When this happens, the star is called a pulsar (denoted PSR). It is estimated that there are as many as a billion neutron stars in the Milky Way, of which about 3000 are pulsars. Pulsars are like enormous flywheels whose pulses are so regular that these stars are considered to be the most precise clocks in the universe. So precise that one tries to exploit this extraordinary characteristic to perform likewise precise measurements that would otherwise be impossible. Some neutron stars have a compact companion, that can either be another neutron star or a white dwarf. In particular, when one of the companions is a pulsar, we speak of a \emph{binary pulsar}. The most famous, and the first to be discovered binary pulsar, known as PSR B1913+16, is a binary star system composed by a pulsar and a companion neutron star. It was discovered in 1974 by Russell A. Hulse and Joseph H. Taylor \cite{hulse} and the careful study of the relative orbital motion of the two stars and, in particular, the observation of the slow decrease of its orbital period, provided the first indirect conclusive evidence of the existence of gravitational waves. In 2003 a double pulsar was discovered \cite{burgay}, namely a system of two companion stars that are both pulsars. It is known as PSR J0737-3039 and it is the only known double pulsar to date. Binary pulsars are rare, but they represent a very important opportunity to test the validity of gravity theories with extreme refinement. These systems can provide verifications of gravity theories complementary to those allowed by the analysis of the gravitational waves emitted in the merging of pairs of compact objects, such as GW170817. Every gravity theory different from Newton's or Einstein's theory foresees deviations about the behaviour of the motion of a binary system predicted by these theories. These deviations are very small, but thanks to the time precisions of binary pulsars they can allow to compare the predictions of general relativity with those of the other modified relativistic theories (RMOND, F(R), Brans--Dicke etc.). And, to date, no deviations from general relativity have been found, which leads to refute many of the alternative theories.

\subsection{Unexpected effects of general relativity}\label{unespected}
There exists a third line of research, even more minority than the one advocating a modified gravity, according to which the phenomena ascribed to dark matter would not only be fictitious, but they would not even be necessary to modify the fundamental laws of gravity, since the required modifications to Newtonian dynamics are already there, though they have not yet been sufficiently analysed in an alternative theory: Einstein's theory of general relativity.\\
Due to the extreme precision with which it has been confirmed, both in observations about the behaviour of binary pulsars \cite{ding}, as well as in the analysis of the gravitational waves originated from mergers of compact objects \cite{ghosh}, it is reasonable to hypothesise that general relativity may not require additional modifications. This is also assumed in dark matter theories.\footnote{In general, these theories presume that, with the exclusion of a few phenomena, the Newtonian approximation to general relativity would be sufficient.} Indeed, general relativity is a much more complex theory than Newton's gravity. Its gravitational field is dynamical, it propagates by waves, and, in contrast with Newton's gravitational potential, which is a scalar, it is described by a multicomponent geometrical tensor field obeying highly nonlinear equations. Usually, in many phenomena associated to dark matter, as in the case of the galaxy rotation curves and of the virial of clusters, one disregards entirely the relativistic effects, and feels safe in simply applying Newton's theory. In this, one is justified by the fact that, in such systems, typical speeds are much smaller than the speed of light, and gravitational fields are very weak due to the low energy density, taking for granted that relativistic corrections are very small compared to Newtonian effects. Over small regions this is certainly true, but on sufficiently large scales the effects of general relativity could be more important than expected, so that it is worth while to investigate if these effects could play a non-negligible role in large scale dynamics even at low speeds and low energy \cite{bg, crosta, gorini, alba}.\\ 
Certain dark matter phenomena have a relativistic nature that can't be ignored, as in the case of gravitational lenses and CMB multipoles. Even for these, however, one normally applies a simplified version of general relativity. This can be understood, though not necessarily justified, since general relativity is such a complicated theory that its equations become utterly difficult to solve, in absence of simplifying assumptions. For example, gravitational lensing is usually calculated by considering the mass concentrated in one point or in a finite number of massive points \cite{narayan}, or by resorting to the simplifying assumption of extended compact lensing bodies of uniform density \cite{slava}. Likewise, in general, the models for the expansion of the universe presuppose the latter to be homogeneous and isotropic at large scales. And so on.\\
In spite of it being a little shared opinion, it makes perfectly sense to investigate in depth how much the exact Einstein equations, and various often neglected general relativistic effects, may affect the dynamics at large scales, independently of the fact whether or not they would lead to a solution of the problem of dark energy. For example, in the same way as matter--energy curves and distort spacetime, the motion of large masses may even, in particular, drag space, and more generally spacetime, along, an effect that, appropriately, has been termed \emph{frame dragging}. A typical example of this phenomenon is the so called Lense--Thirring effect \cite{thirring 1917, thirring 1918, lense}, whereby matter-energy spin and/or orbital angular momentum couple and hence precess about a common axis. This effect is akin to the Larmor precession in electrodynamics \cite{jackson}. It is due to the fact that mass currents generate a type of gravitational field that is the analogue of the magnetic field generated by an electric current, and that is appropriately called a gravitomagnetic field. For example, the effect has been observed both on the Lageos satellites and on gyroscopes orbiting the spinning Earth \cite{ciufolini, ciufoloni, everitt}, as well as in binary pulsars, where the spin axes of the orbiting stars precess about a common axis \cite{kramer, breton}. The frame dragging phenomenon is totally non Newtonian. It is small locally, but its building-up on large scales may accumulate in important deviations of the global dynamics compared to the Newtonian description. For example, in the case of a disc galaxy, which may be composed by a few hundred billion stars, though its local behaviour is Newtonian, its angular momentum is so high that it might be responsible of large frame dragging effects that may, at least partly, account for its missing mass.\\
Furthermore, by the fact that, unlike the nonrelativistic description, general relativity predicts that gravity propagates at finite speed, one might expect that retardation effects in place of instantaneous transmission of the potential may make themselves felt in a non negligible way, particularly on scales from thousands of light years and more. For example, the mass variations of galaxies in time, due to the interactions with satellite galaxies, may give rise to retarded potentials that may contribute to the estimate of the content of dark matter. The same may take place as a consequence of the action of far away clusters in relative motion with respect to the galaxy under observation.\\
Finally, it is crucial to face the problem of the identification of the observer compared to the coordinate frames in general relativity \cite{lusanna, lusanna 1, lusanna 2}. This is a very delicate issue, that calls for an attention much bigger than the one that, with rare exceptions, it has received by the scientific community.

\section{Future perspectives}\label{DM:2.4}
Many future enterprises, both on Earth (Earth--based telescopes), as well as in space (space--based telescopes), are active or or in construction with, among other things, the purpose of trying to clarify the nature and properties of both dark energy and dark matter. We have listed and briefly described them in the conclusive Section 4 of the Chapter on dark energy. Therefore, we refer to this Chapter for the relevant information. 




\begin{thebibliography}{99}

\bibitem{url} \url{https://link.springer.com/book/10.1007/978-3-031-61187-2}

\bibitem{KGB} K. G. Begeman, ``HI rotation curves of spiral galaxies,'' PhD Thesis 1987, https://research.rug.nl/en/publications/hi-rotation-curves-of-spiral-galaxies
    
\bibitem{macdougal 2012} Douglas W. MacDougal (2012) ``Newton's Gravity: An Introductory Guide to the Mechanics of the Universe'', Springer New York 2012, DOI:10.1007 978-1-4614-5444-1

\bibitem{zwicky1} F. Zwicky (1933) ``Die Rotverschiebung von extragalaktischen Nebeln,'' in Helvetica Physica Acta, vol. 6, 110 --127

\bibitem{zwicky2} F. Zwicky (1937) ``On the Masses of Nebulae and of Clusters of Nebulae,'' in Astrophysical Journal, vol. 86, 217

\bibitem{eddington 19} F. W. Dyson, A. S. Eddington, C. Davidson (1920) ``A determination of the deflection of light by the Sun's gravitational field, from observations made at the total eclipse of 29 May 1919,'' Philosophical Transactions of the Royal Society, 220A (571–581): 291–333.

\bibitem{gravlens} D. Walsh, R. F. Carswell, R. J. Weymann (1979) ``0957 + 561 A, B: twin quasistellar objects or gravitational lens?,'' Nature. 279 (5712): 381--384.

\bibitem{brandemberger} R.~H.~Brandenberger (2001) ``Frontiers of inflationary cosmology,'' Braz. J. Phys. \textbf{31}, 131-146

\bibitem{coley} A.~A.~Coley (2023) ``The JWST and standard cosmology,'' [arXiv:2312.14738 [gr-qc]].

\bibitem{abdalla} Elcio Abdalla, Guillermo Franco Abell\`an {\it et al.} (2022) ``Cosmology Intertwined: A Review of the Particle Physics, Astrophysics, and Cosmology Associated with the Cosmological Tensions and Anomalies,'' Journal of High Energy Astrophysics. 34: 49.

\bibitem{el gordo} F.~Menanteau, J.~P.~Hughes, C.~Sifon, \textit{et al.} (2012) ``The Atacama Cosmology Telescope: ACT-CL J0102-4215 'El Gordo,' a Massive Merging Cluster at Redshift 0.87,'' Astrophys. J. \textbf{748}, 7

\bibitem{smoot} G.~F.~Smoot (2007) ``CMB Anisotropies: Their Discovery and Utilization,'' Nuovo Cim. B \textbf{122}: 1339-1351

\bibitem{smoot1} G. F. Smoot, C. L. Bennett, A. Kogut, \textit{et al.} (1992) ``Structure in the COBE Differential Microwave Radiometer First-Year Maps,'' Astrophysical Journal Letters v.396, p.L1.

\bibitem{komatsu} E. Komatsu {\it et al.} (2009) ``FIVE-YEAR WILKINSON MICROWAVE ANISOTROPY PROBE$^*$ OBSERVATIONS: COSMOLOGICAL INTERPRETATION,'' ApJS \textbf{180} 330

\bibitem{planck} Planck Collaboration (2020) ``{\it Planck} 2018 results,'' A\&A 641, A1

\bibitem{legendre} Harry Hochstadt (2012) ``The Functions of Mathematical Physics,'' Dover Publications.

\bibitem{sunyaev} R. A. Sunyaev, Ya. B. Zel'dovich (1970) ``Small-Scale Fluctuations of Relic Radiation,'' Astrophysics and Space Science. 7 (1): 3--19.

\bibitem{sunyaev1} R. A. Sunyaev, Ya. B. Zel'dovich (1980) ``Microwave background radiation as a probe of the contemporary structure and history of the universe,'' Annual Review of Astronomy and Astrophysics. 18 (1): 537--560

\bibitem{standard model} W. N. Cottingham and D. A. Greenwood (2007) ``AN INTRODUCTION TO THE STANDARD MODEL OF PARTICLE PHYSICS,''  Cambridge University Press,  Cambridge UK 

\bibitem{sterile neutrinos} Basudeb Dasgupta, Joachim Kopp (2021) ``Sterile neutrinos,'' Physics Reports, 928, pages 1--63

\bibitem{massimi} Michela Massimi, (2005) ``Pauli's Exclusion Principle,'' Cambridge University Press, Cambridge

\bibitem{chandrasekhar mass} S. Chandrasekhar (1931) ``The Maximum Mass of Ideal White Dwarfs,'' Astrophysical Journal 74: 81--82.

\bibitem{explosions} S.~W.~Hawking (1974) ``Black hole explosions,'' Nature \textbf{248}: 30--31

\bibitem{1976} S.~W.~Hawking (1975) ``Particle Creation by Black Holes,'' Commun. Math. Phys. \textbf{43}, 199--220; Erratum: Commun. Math. Phys. \textbf{46} (1976), 206

\bibitem{pavel} Pavel Kroupa (2023), private communication.

\bibitem{novikov} Ya. B. Zel'dovitch, I. Novikov (1966) ``The Hypothesis of Cores Retarded During Expansion and the Hot Cosmological MOdel,'' Soviet Astronomy. 10 (4): 602--603.

\bibitem{hawking} S. Hawking (1971) ``Gravitationally collapsed objects of very low mass,'' Mon. Not. R. Astron. Soc. 152: 75.

\bibitem{chadwick} James Chadwick (1932) ``The Existence of a Neutron,'' Proceedings of the Royal Society of London, Series A, 136, No. 830, pp. 692--708

\bibitem{cowan reines} C. L. Cowan Jr., F. Reines, F. B. Harrison, H. W. Kruse, A. D. McGuire (1956) ``Detection of the Free Neutrino: a Confirmation,'' Science, 124 (3212): 103--104

\bibitem{weinberg super} S.~Weinberg (2013) ``The quantum theory of fields. Vol. 3: Supersymmetry,'' Cambridge University Press, Cambridge

\bibitem{lee yang} T. D. Lee and C. N. Yang (1956) ``Question of Parity Conservation in Weak Interactions,'' Phys. Rev. 104, 254; Erratum (1957) Phys. Rev. 106, 1371

\bibitem{noether} E. Noether (1918) ``Invariante Variationsprobleme,'' Nachrichten von der Gesellschaft der Wissenschaften zu Göttingen. Mathematisch-Physikalische Klasse. 1918: 235--257

\bibitem{wu} Jianxiong Wu, E. Ambler, R. W. Hayward (1957) ``Experimental Test of Parity Conservation in Beta Decay,'' Physical Review, vol. 105, n. 4: 1413--1415,

\bibitem{ioffe} B. L. Ioffe, L. B. Okun, A. P. Rudik (1957) ``The Problem of Parity Non-conservation in Weak Interactions,'' Journal of Experimental and Theoretical Physics. 32: 328--330.

\bibitem{landau}  L. D. Landau (1957) ``On the conservation laws for weak interactions,'' Nuclear Physics 3 (1): 127--131

\bibitem{cronin} J. H. Christenson, J. W. Cronin, V. L. Fitch and R. Turlay (1964) ``Evidence for the $2\pi$ Decay of the $K_2^0$ Meson," Phys. Rev. Lett. 13, 138 

\bibitem{peccei} R. D. Peccei, H. R. Quinn (1977) ``CP Conservation in the Presence of Pseudoparticles,'' Physical Review Letters 38 (25): 1440--1443

\bibitem{quinn}  R. D. Peccei, H. R. Quinn (1977) ``Constraints imposed by CP conservation in the presence of pseudoparticles,'' Physical Review D 16 (6): 1791--1797

\bibitem{wilczek} Frank Wilczek (1978) ``Problem of Strong P and T Invariance in the Presence of Instantons,'' Physical Review Letters 40 (5): 279--282

\bibitem{axion} Steven Weinberg (1978) ``A New Light Boson?'' Physical Review Letters 40 (4): 223--226.

\bibitem{shallow} Lanre Akinyemi (2023) ``Shallow ocean soliton and localized waves in extended $(2+1)$-dimensional nonlinear evolution equations,'' Physics Letters A,
Volume 463: 128668

\bibitem{primakov} H. Primakoff (1951) ``Photo-Production of Neutral Mesons in Nuclear Electric Fields and the Mean Life of the Neutral Meson,'' Phys. Rev. \textbf{81}, 899

\bibitem{amrut 2023} Amruth, A., Broadhurst, T., Lim, J. {\it et al.} (2023) ``Einstein rings modulated by wavelike dark matter from anomalies in gravitationally lensed images.'' Nat Astron 7, 736--747

\bibitem{bird 2023} Simeon Bird {\it et al.} (2023) ``Enhanced Early Galaxy Formation in JWST from Axion Dark Matter?'' [ArXiv:2307.10302v1[hep--ph]]  

\bibitem{jwst} A.~A.~Coley (2023) ``The JWST and standard cosmology,'' [arXiv:2312.14738 [gr-qc]]

\bibitem{z11} K. Glazebrook, T. Nanayakkara, C., Schreiber, {\it et al.} (2024) ``A massive galaxy that formed its stars at $z\sim11$,'' Nature (2024),  https://doi.org/10.1038/s41586-024-07191-9 

\bibitem{mack} K.~J.~Mack, (2011) ``Axions, Inflation and the Anthropic Principle,'' JCAP \textbf{07}: 021

\bibitem{lella} Federico Lelli, Stacy S. McGaugh, and James M. Schombert (2016) ``SPARC: MASS MODELS FOR 175 DISK GALAXIES WITH SPITZER PHOTOMETRY
AND ACCURATE ROTATION CURVES,'' The Astronomical Journal, 152:157

\bibitem{milgrom 1} M.~Milgrom (1983) ``A Modification of the Newtonian dynamics as a possible alternative to the hidden mass hypothesis,'' Astrophys. J. \textbf{270}, 365--370

\bibitem{milgrom 2} M.~Milgrom (1983) ``A Modification of the Newtonian dynamics: Implications for galaxies,'' Astrophys. J. \textbf{270}, 371--383

\bibitem{milgrom 3} M.~Milgrom (1983) ``A modification of the Newtonian dynamics: implications for galaxy systems,'' Astrophys. J. \textbf{270}, 384 --389

\bibitem{freeman 1970} K. C. Freeman (1970) ``On the disks of spiral and S0 galaxies,'' Astrophys. J. \textbf{160}, 811

\bibitem{fish 1964} R. A. Fish (1964) ``A Mass-Potential Relationship in Elliptical Galaxies and Some Inferences Concerning
the Formation and Evolution of Galaxies.'' Astrophys. J. \textbf{139}, 284

\bibitem{udg} J.~Koda, M.~Yagi, H.~Yamanoi and Y.~Komiyama (2015) ``Approximately A Thousand Ultra Diffuse Galaxies in the Coma cluster,'' Astrophys. J. Lett. \textbf{807} no.1, L2

\bibitem{asencio 2022} Elena Asencio, Indranil Banik, Steffen Mieske, {\it et al.} (2022) ``The distribution and morphologies of Fornax Cluster dwarf galaxies suggest they lack dark matter,'' MNRAS, 515, Issue 2: 2981--3013

\bibitem{chae 2020} Kyu-Hyun Chae {\it et al.} (2020) ``Testing the Strong Equivalence Principle: Detection of the External Field Effect in Rotationally Supported Galaxies,'' Astrophys. J. \textbf{904}, 51

\bibitem{angus 2007} Garry W. Angus {\it et al.} (2007) ``On the Proof of Dark Matter, the Law of Gravity, and the Mass of Neutrinos,'' Astrophys. J. \textbf{654}, L13

\bibitem{sanders 2007} R. H. Sanders (2007) ``Neutrinos as cluster dark matter,'' MNRAS, 380, Issue 1: 331--338

\bibitem{nieuwenhuizen 2016} Theodorus Maria Nieuwenhuizen (2017) ``How Zwicky already ruled out modified gravity theories without dark matter,'' Fortschritte der Physik 65(6--8)

\bibitem{aguirre 2001} Anthony Aguirre {\it et al.} (2001) ``Problems for Modified Newtonian Dynamics in Clusters and the Ly$\alpha$ Forest?'' Astrophys. J. \textbf{561}, 550

\bibitem{van dokkum 2022} P. van Dokkum, Z. Shen, M. A. Keim, {\it et al.} (2022) ``A trail of dark-matter-free galaxies from a bullet-dwarf collision.'' Nature 605, 435--439

\bibitem{silk 2029} Joseph Silk (2019) ``Ultra-diffuse galaxies without dark matter,'' MNRAS, 488, Issue 1: L24--L28 

\bibitem{mancera 2022} Pavel E. Mancera Pi\~na, Filippo Fraternali, {\it et al.} (2022) ``No need for dark matter: resolved kinematics of the ultra-diffuse galaxy AGC 114905,'' MNRAS, 512, Issue 3: 3230--3242,

\bibitem{moreno 2022} J. Moreno, S. Danieli, J. S. Bullock, {\it et al.} (2022) ``Galaxies lacking dark matter produced by close encounters in a cosmological simulation.'' Nat. Astron. 6, 496--502

\bibitem{shin} Eun-Jin Shin {\it et al.} (2020) ``Dark matter deficient galaxies produced via high velocity galaxy collisions in high-resolution numerical simulations,'' ApJ 899:25 

\bibitem{muller 2019} Oliver M\"uller, Benoit Famaey and Hongsheng Zhao (2019) ``Predicted MOND velocity dispersions for a catalog of ultra--diffuse galaxies in group environments,'' A\&A 623, A36

\bibitem{banik 2022} Indranil Banik, Srikanth T Nagesh, Hosein Haghi, {\it et al.} (2022) ``Overestimated inclinations of Milgromian disc galaxies: the case of the ultradiffuse galaxy AGC 114905,'' MNRAS, 513, Issue 3: 3541--3548

\bibitem{angus 2006} G. W. Angus, B. Famaey, H. S. Zhao (2006) ``Can MOND take a bullet? Analytical comparisons of three versions of MOND beyond spherical symmetry,'' MNRAS, 371, Issue 1: 138--146

\bibitem{rodrigues 2018} D. C. Rodrigues, V. Marra, A. del Popolo, {\it et al.} (2018) ``Absence of a fundamental acceleration scale in galaxies.'' Nat Astron 2, 668--672

\bibitem{kroupa 2018} P. Kroupa, I. Banik, H. Haghi, {\it et al.} (2018) ``A common Milgromian acceleration scale in nature.'' Nat. Astron. 2, 925--926

\bibitem{mcgaugh 2018} S. S. McGaugh, P. Li, F. Lelli, {\it et al.} (2018) ``Presence of a fundamental acceleration scale in galaxies.'' Nat. Astron. 2, 924

\bibitem{comeron 2023} Sebastien Comer\'on {\it et al.} (2023) ``The massive relic galaxy NGC 1277 is dark matter deficient. From dynamical models of integral--field stellar kinematics and to five effective radii, A\&A 675, A143

\bibitem{mah} S. S. McGaugh in https://tritonstation.com/2023/08/02/is-ngc-1277-a-problem-for-mond/ 

\bibitem{banik 2024} Indranil Banik, Charalambos Pittordis, Will Sutherland, {\it et al.} (2024) ``Strong constraints on the gravitational law from Gaia DR3 wide binaries,'' MNRAS, 527, Issue 3: 4573–4615

\bibitem{chae 2023} Kyu-Hyun Chae (2023) ``Breakdown of the Newton--Einstein Standard Gravity at Low Acceleration in Internal Dynamics of Wide Binary Stars,'' Astrophys. J. \textbf{952}, 128

\bibitem{hernandez 2023} X. Hernandez (2023) ``Internal kinematics of Gaia DR3 wide binaries: anomalous behaviour in the low acceleration regime,'' MNRAS, 525, Issue 1: 1401--1415

\bibitem{bekenstein 2004} Jacob D. Bekenstein (2004) ``Relativistic gravitation theory for the modified Newtonian dynamics paradigm,'' Phys. Rev. D 70, 083509; (2005) Erratum Phys. Rev. D 71, 069901

\bibitem{clifton 2012} Timothy Clifton, Pedro G. Ferreira, Antonio Padilla, Constantinos Skordis (2012) ``Modified gravity and cosmology,'' Physics Reports, 513, Issues 1--3: 1--189,

\bibitem{skordis 2006} C. Skordis, D. F. Mota, P. G. Ferreira, and C. B{\ae}hm (2006) ``Large Scale Structure in Bekenstein’s Theory of Relativistic Modified Newtonian Dynamics,'' Phys. Rev. Lett. 96, 011301

\bibitem{slosar 2005} An\v{z}e Slosar, Alessandro Melchiorri, and Joseph I. Silk (2005) ``Test of modified Newtonian dynamics with recent Boomerang data,'' Phys. Rev. D \textbf{72}, 101301

\bibitem{mavromatos 2009} Nick E. Mavromatos, Mairi Sakellariadou, and Muhammad Furqaan Yusaf (2009) ``Can the relativistic field theory version of modified Newtonian dynamics avoid dark matter on galactic scales?'' Phys. Rev. D \textbf{79}, 081301

\bibitem{seifert 2007} Michael D. Seifert (2007) ``Stability of spherically symmetric solutions in modified theories of gravity,'' Phys. Rev. D 76, 064002

\bibitem{sanders 2005} R. H. Sanders (2005) ``A tensor—vector—scalar framework for modified dynamics and cosmic dark matter,'' MNRAS, 363, Issue 2: 459--468

\bibitem{moffat 2006} J. W. Moffat (2006) ``Scalar--tensor--vector gravity theory,'' JCAP03(2006)004

\bibitem{skordis 2021} Constantinos Skordis and Tom Z\l o\'snik (2021) ``New Relativistic Theory for Modified Newtonian Dynamics,'' Phys. Rev. Lett. 127, 161302

\bibitem{buchdahl} H. A. Buchdahl (1970) ``Non-linear Lagrangians and cosmological theory,'' MNRAS 150: 1--8

\bibitem{brans} C. H. Brans, R. H. Dicke (1961) ``Mach's Principle and a Relativistic Theory of Gravitation,'' Physical Review. 124 (3): 925--935

\bibitem{verlinde 2011} E. Verlinde (2011) ``On the origin of gravity and the laws of Newton.'' J. High Energ. Phys. 2011, 29

\bibitem{schlatter 2023} A. Schlatter and R. E. Kastner (2023) ``Gravity from transactions: fulfilling the entropic gravity program,'' J. Phys. Commun. 7:065009

\bibitem{Einstein 1915} A. Einstein (1915) ``Zur Allgemeinen Relativit\"atstheorie, Sitzungsber,'' Preuss. Akad. Wiss. Berlin (Math. Phys.) \textbf{1915}, 778-786

\bibitem{gw150914} B. P. Abbott {\it et al. (LIGO Scientific Collaboration and Virgo Collaboration)} (2016) ``Observation of Gravitational Waves from a Binary Black Hole Merger,'' Phys. Rev. Lett. 116, 061102 

\bibitem{gw170817} B.~P.~Abbott \textit{et al. (LIGO Scientific and Virgo)} (2017) ``GW170817: Observation of Gravitational Waves from a Binary Neutron Star Inspiral,'' Phys. Rev. Lett. \textbf{119} no.16, 161101

\bibitem{skordis 2019} Constantinos Skordis and Tom Z\l o\'snik (2019) ``Gravitational alternatives to dark matter with tensor mode speed equaling the speed of light,'' Phys. Rev. D 100, 104013

\bibitem{hulse} R. A. Hulse, J. H. Taylor (1974) ``A High-Sensitivity Pulsar Survey,'' Astrophys. J. 191 L59

\bibitem{burgay} M. Burgay, N. d'Amico, {\it et al.} (2003) ``An increased estimate of the merger rate of double neutron stars from observations of a highly relativistic system,'' Nature. 426 (6966): 531--533

\bibitem{ding} Hao Ding {\it et al.} (2021) ``The Orbital-decay Test of General Relativity to the $2\%$ Level with 6 yr VLBA Astrometry of the Double Neutron Star PSR J1537+1155,''  ApJL \textbf{921} L19

\bibitem{ghosh} A.~Ghosh, N.~K.~Johnson-Mcdaniel, A.~Ghosh, {\it et al.} (2018) ``Testing general relativity using gravitational wave signals from the inspiral, merger and ringdown of binary black holes,'' Class. Quant. Grav. \textbf{35} no.1, 014002

\bibitem{bg} H.~Balasin and D.~Grumiller (2008) ``Non-Newtonian behavior in weak field general relativity for extended rotating sources,'' Int. J. Mod. Phys. D \textbf{17}, 475--488

\bibitem{crosta}  W.~Beordo, M.~Crosta, M.~G.~Lattanzi, P.~Re Fiorentin and A.~Spagna (2024) ``Geometry-driven and dark-matter-sustained Milky Way rotation curves with Gaia DR3,''
Mon. Not. Roy. Astron. Soc. \textbf{529} (2024) no.4, 4681-- 4698

\bibitem{gorini} D.~Astesiano, S.~L.~Cacciatori, V.~Gorini and F.~Re  (2022) ``Towards a full general relativistic approach to galaxies,'' Eur. Phys. J. C \textbf{82}, no.6, 554

\bibitem{alba} D.~Alba and L.~Lusanna (2012) ``The Einstein-Maxwell-Particle System in the York Canonical Basis of ADM Tetrad Gravity: III) The Post-Minkowskian N-Body Problem, its Post-Newtonian Limit in Non-Harmonic 3-Orthogonal Gauges and Dark Matter as an Inertial Effect,'' Can. J. Phys. \textbf{90}, 1131-1178

\bibitem{narayan} R.~Narayan and S.~Wallington (1992) ``Introduction to basic concepts of gravitational lensing,'' Lect. Notes Phys. \textbf{406}, 12--26

\bibitem{slava} Slava G. Turyshev and Viktor T. Toth (2022) ``Recovering the mass distribution of an extended gravitational lens,'' MNRAS 513(4), 5355--5376

\bibitem{thirring 1917} H. Thirring (1917) ``Wirkung rotierender Massen,'' Österr. Zentralbibl. Wien
 
\bibitem{thirring 1918} H. Thirring (1918) ``\"Uber die Wirkung rotierender ferner Massen in der Einsteinschen Gravitationstheorie,'' Phys. Zs. 19, 33 [English translation in Gen. Rel. Grav. 16 (1984), 712]

\bibitem{lense} J. Lense, H. Thirring (1918) ``Ueber den Einfluss der Eigenrotation der Zentralkoerper auf die Bewegung der Planeten und Monde nach der Einsteinschen Gravitationstheorie,'' Phys. Zs. 19:156--163 (1918). (Translation in Gen. Rel. Grav. 16:727 (1984))

\bibitem{jackson} J. D. Jackson (1999) ``Classical Electrodynamics,'' 3rd edition, Wiley, New York

\bibitem{ciufolini} I. Ciufolini, D. Lucchesi, F. Vespe, F. Chieppa (1997) ``Detection of Lense–Thirring Effect Due to Earth's Spin,'' [arXiv:gr-qc/9704065]

\bibitem{ciufoloni} I. Ciufolini {\it et al.} (1997) TEST OF LENS-THIRRING ORBITAL SHIFT DUE TO SPIN, Classical and quantum gravity, 14(10), pp. 2701-2726

\bibitem{everitt} C. W. F. Everitt {\it et al.} (2011) ``Gravity Probe B: Final Results of a Space Experiment to Test General Relativity,'' Phys. Rev. Lett. 106: 221101

\bibitem{kramer} Michael Kramer (1998) ``Determination of the Geometry of the PSR B1913+16 System by Geodetic Precession,''
THE ASTROPHYSICAL JOURNAL, 509:856--860

\bibitem{breton} R.~P.~Breton, V.~M.~Kaspi, M.~Kramer, {\it et al.} (2008) ``Relativistic Spin Precession in the Double Pulsar,'' Science \textbf{321}, 104--107

\bibitem{lusanna} L.~Lusanna (2010) ``Post-Minkowskian Gravity: Dark Matter as a Relativistic Inertial Effect?,'' J. Phys. Conf. Ser. \textbf{222}, 012016

\bibitem{lusanna 1} L.~Lusanna (2011) ``Canonical Gravity and Relativistic Metrology: from Clock Synchronization to Dark Matter as a Relativistic Inertial Effect,'' [arXiv:1108.3224 [gr-qc]]

\bibitem{lusanna 2} L.~Lusanna (2017) ``Dark matter: a problem in relativistic metrology?,'' J. Phys. Conf. Ser. \textbf{845}, no.1, 012007











    
\end{thebibliography}
\end{document}